\documentclass
[superscriptaddress,secnumarabic,amssymb,amsmath,nobibnotes,aps,prd,showkeys,showpacs,nofootinbib,onecolumn]{revtex4}%
\usepackage{pdflscape}
\usepackage{graphicx}
\usepackage{subfigure}
\usepackage{epstopdf}
\usepackage{epsf}
\usepackage{bm}
\usepackage{amsmath}
\usepackage{amsfonts}
\usepackage{amssymb}
\usepackage[T1]{fontenc}
\usepackage[colorlinks=true,             linkcolor=blue, urlcolor=blue,
citecolor=blue]{hyperref}
\usepackage[colorinlistoftodos]{todonotes}%
\providecommand{\U}[1]{\protect\rule{.1in}{.1in}}

\newcommand{\be}{\begin{equation}}
\newcommand{\ee}{\end{equation}}

\newcommand{\mincir}{\raise
-3.truept\hbox{\rlap{\hbox{$\sim$}}\raise4.truept\hbox{$<$}\ }}
\newcommand{\magcir}{\raise
-3.truept\hbox{\rlap{\hbox{$\sim$}}\raise4.truept\hbox{$>$}\ }}

\begin{document}
\title{Exact Kantowski-Sachs spacetimes in Einstein-Aether Scalar field theory}
\author{Genly Leon}
\email{genly.leon@ucn.cl}
\affiliation{Departamento de Matem\'{a}ticas, Universidad Cat\'{o}lica del Norte, Avda.
Angamos 0610, Casilla 1280 Antofagasta, Chile.}
\author{Andronikos Paliathanasis}
\email{anpaliat@phys.uoa.gr}
\affiliation{Institute of Systems Science, Durban University of Technology, Durban 4000,
South Africa}
\affiliation{Instituto de Ciencias F\'{\i}sicas y Matem\'{a}ticas, Universidad Austral de
Chile, Valdivia 5090000, Chile}
\author{N. Dimakis}
\email{nsdimakis@scu.edu.cn ; nsdimakis@gmail.com}
\affiliation{Center for Theoretical Physics, College of Physics Sichuan University, Chengdu
610064, China}

\begin{abstract}
Exact and analytic solutions in Einstein-Aether scalar field theory with
Kantowski-Sachs background space are determined. The theory of point
symmetries is applied to determine the functional form of the unknown
functions which defines the gravitational model. Conservation laws are applied
to reduce the order of the field equations and write the analytic solution.
Moreover, in order to understand the physical behaviour of the cosmological
model a detailed analysis of the asymptotic behaviour for the gravitational
field equations is performed.

\end{abstract}
\date{\today}
\maketitle

\section{Introduction}

The detailed analysis of the cosmic microwave background (CMB) shows the
existence of anisotropies which are small enough to suggest that anisotropic
models of spacetimes become isotropic by evolving in time \cite{Mis69, szydl,
russ}. According to this scenario, the mechanism which explains the
isotropization of the observable universe today is called inflation
\cite{guth}. Inflation occurs when the potential part of a scalar field, known
as inflaton, dominates in the gravitational field equations and drives the
dynamics. Because anisotropic spacetimes describe the pre-inflationary era,
their analysis in the presence of scalar fields is of special interest for the
study of the early universe.

The main class of cosmological models which describe spatially homogeneous and
(in general) anisotropic spacetimes consists of the Bianchi cosmologies. There
are various studies in the literature with the existence of scalar fields in
Bianchi spacetimes. Some exact anisotropic spacetimes are determined in
\cite{b1} where it is found that an exponential scalar field provides
power-law scale factors. The asymptotic behaviour of the dynamics for the
Bianchi I model with exponential potential is studied in \cite{b2} while the
analytic solution of the later model is found in \cite{b3}. Some other studies
on anisotropic universes are presented in \cite{b4,b5,b6,b7,b8} and references therein.

In this work we are interested in the Lorentz violating inflationary model
proposed in \cite{kanno}. That model belongs to the family of Einstein-Aether
scalar field theory \cite{DJ} where the scalar field is coupled to the Aether
field and there is an interaction between the two \cite{kanno}. In particular,
the coefficient components which define the Aether Action Integral are assumed
to be functions of the scalar field. In this case the field equations provide
two inflationary stages, the Lorentz-violating stage and the standard
slow-roll. In the Lorentz-violating state, the universe expands as an exact de
Sitter spacetime, although the inflaton field is rolling down the potential.

In the context of exact and analytic solutions for the Einstein-Aether scalar
field theory there are very few studies in the literature \cite{eas1,eas2}. As
far as the inflationary model of Kanno and Soda \cite{kanno} is concerned, in
\cite{ea1,ea2} the unknown functions of the model were determined in the cases
of a Friedmann--Lema\^{\i}tre--Robertson--Walker or Bianchi I background
spaces so as for the gravitational field equations to admit conservation laws
and the resulting gravitational system to be Liouville integrable. In this
work, we extend the latter analysis by considering a Kantowski-Sachs
background space \cite{KS1}. The Kantowski-Sachs spacetime is a locally
rotational spacetime which admits the isometry group $\mathbb{R}\times SO(3)$,
which does not act simply transitively on the spacetime, nor does its three
dimensional subgroup have a simple transitive action on some spacelike
hypersurface. Hence - even though the model is spatially homogeneous - it does
not belong to the Bianchi classification. An interesting characteristic of the
Kantowski-Sachs model is that in the limit of isotropization the dynamics of
the scale factor resemble those of a closed
Friedmann--Lema\^{\i}tre--Robertson--Walker space-time.

Integrability is an important property in all areas of mathematical
physics. Nowadays, numerical techniques are mainly applied to solve nonlinear
dynamical systems; thus, the demonstration that a system is integrable
indicates that the numerical solutions correspond to actual solutions of the
dynamical system. Sometimes, integrable dynamical systems can be described by
closed-form functions, which means that closed-form analytic or exact
solutions exist (which is usually referred to as Liouville integrability). Although, an
integrable dynamical system may not describe completely a real physical
system, that is, physical observations, it can still be used as a toy model in order to
study the viability of the given theory. In gravitational theory, the field
equations form a nonlinear dynamical system where various techniques can be
applied to investigate if the field equations possess the integrability
property. In \cite{ea1} the field equations were solved for the
Einstein-aether theory given by Kanno and Soda \cite{kanno} in the context of a
flat FLRW metric. The minisuperspace approach was used and five
classes of scalar field potentials were found for a quadratic coupling between the scalar
field and the Aether field, using as mathematical criteria the existence of
solutions for Liouville integrable equations. Following this line, the
analysis was extended to Bianchi I models (which is the natural extension of
flat FLRW to the anisotropic set up) in \cite{ea2}. Additionally to the
integrability of the equations in the sense of Liouville, the stability of the
equilibrium points was discussed, and the evolution of the anisotropies was
studied in \cite{ea2}. 

Using the Hubble-normalized variables, and combining with alternative
dimensionless variables (which lead to the evolution of anisotropies with
local and with Poincare variables) it was concluded that the isotropic
spatially flat FLRW spacetime is a future attractor for the physical space.
However, Kasner-like anisotropic solutions are also allowed by the theory.
It is well-know in the GR case that for Bianchi I and Bianchi III, the
Hubble parameter $H$ is always monotonic and the anisotropy decays in time for
$H>0$. Therefore, isotropization occurs \cite{nns1}. However, for
Kantowski-Sachs, as well as for closed FLRW, the Hubble parameter is not
guaranteed to be monotonic, and anisotropies would increase rather than
vanish (see \cite{Byland:1998gx}, \cite{Fadragas:2013ina} and references
therein). For a perfect fluid in Kantowski-Sachs Einstein-aether theory
without scalar field \cite{Latta:2016jix}, solutions were found that either
expand from or contract to anisotropic states. A partial proof of this was
given in \cite{Coley:2015qqa}. These solutions are a non-trivial consequence
of the presence of a non-zero Lorentz-violating vector field. Kantowski-Sachs
metrics also admit Einstein's static solution. These are crucial
differences with Bianchi I and Bianchi III spacetimes, which makes the analysis of the
Kantowski-Sachs case worth it in the context of  the Einstein-aether theory given by Kanno and Soda
\cite{kanno}. Therefore, the analysis in this paper is a continuation of
papers \cite{ea1} and \cite{ea2}, by considering Kantowski-Sachs in the
Einstein-aether theory given by \cite{kanno}.

The approach that is followed in the determination of solutions for the field
equations is based in the determination of conservation laws for the field
equations. A main property of that specific gravitational model is that the
gravitational field equations admit a minisuperspace and can be derived by the
variation of a point-like Lagrangian. The existence of the latter is essential
because techniques from Analytic Mechanics can be applied \cite{mech3,mech4}
while at the same time it can be used as the base in a quantization process of
the theory, for instance see \cite{eaqm,qm1,qm2}. The approach that we apply
for the determination of conservation laws is that of Lie's theory and in
particular Noether's theorem. This methodology is widely utilized in various
cosmological models with interesting results
\cite{mech1,mech2,ns1,ns2,ns3,ns4,ns5,ns6,ns7,ns8,ns9}.

Kantowski-Sachs have been widely studied in the literature. In the case where
a perfect fluid is introduced the spacetime is egotistically incomplete
\cite{Collins:1977fg}. The case of cosmological constant was studied in
\cite{Weber:1984xh}. On the other hand, an exact solution of field equations
for Kantowski-Sachs background space with cosmological constant was found in
\cite{Gron:1986ua}. For some other studies we refer the reader in
\cite{LorenzPetzold:1985jm,Solomons:2001ef,Jamal:2017cut,Zubair:2016ccy,Alvarenga:2015jaa,Barrow:1996gx,Barrow:2018zav,Byland:1998gx,Calogero:2009mi,Camci:2016yed,Carr:1999qr,deCesare:2020swb,Clancy:1998ka}
and references therein. In the case of Einstein-aether theory in \cite{in1}
the authors presented a generic static spherical symmetric solution, where it
has been shown that the Schwarzschild spacetime is recovered. In addition, the
dynamics of spatially homogeneous Einstein-aether cosmological models with a
scalar field possessing a generalized harmonic potential, in which the scalar
field is coupled to the aether field expansion and shear scalars, are studied
in \cite{col1,col2,Latta:2016jix,Coley:2019tyx,Leon:2019jnu}.

The plan of the paper is as follows: In Section 2, the cosmological model
under consideration is defined and the field equations and the point-like
Lagrangian is presented. The new exact and analytic solutions of the
gravitational field equations for Kantowski-Sachs background space are
presented in Section 3. In Section 4, the asymptotic behaviour of the field
equations is analyzed, which allows to understand the dynamics and the
evolution of the cosmological solutions of the previous section. Finally, in
Section 5, the results are discussed and the conclusions are drawn.

\section{Field equations}

\label{sec2}

The Einstein-Scalar field model proposed by Kanno and Soda \cite{kanno} in
which the Aether coefficients are functions of the scalar field is considered,
that is, the gravitational Action Integral is
\begin{equation}
S=\int dx^{4}\sqrt{-g}\left(  \frac{R}{2}-\frac{1}{2}g^{\mu\nu}\phi_{;\mu}%
\phi_{;\nu}-V\left(  \phi\right)  \right)  -S_{Aether}, \label{ac.01}%
\end{equation}
where $S_{Aether}$ is the Action Integral for the Aether field, defined as
\begin{align}
S_{Aether}  &  =\int dx^{4}\sqrt{-g}\left(  \beta_{1}\left(  \phi\right)
u^{\nu;\mu}u_{\nu;\mu}+\beta_{2}\left(  \phi\right)  \left(  g^{\mu\nu}%
u_{\mu;\nu}\right)  ^{2}\right)  +\nonumber\\
&  +\int dx^{4}\sqrt{-g}\left(  \beta_{3}\left(  \phi\right)  u^{\nu;\mu
}u_{\mu;\nu}+\beta_{4}\left(  \phi\right)  u^{\mu}u^{\nu}u_{\;;\mu}^{\alpha
}u_{\alpha;\nu}-\lambda\left(  u^{\mu}u_{\nu}+1\right)  \right)  ,
\label{ac.02}%
\end{align}
and $\lambda$ is a Lagrange multiplier which ensures the unitarity of the
Aether field $u^{\mu}.~$Functions $\beta_{1},~\beta_{2},~\beta_{3}~$and
$\beta_{4}$ define the coupling between the aether field and the gravitational
field. In the original definition of Einstein-aether theory, coefficients
$\beta_{1},~\beta_{2},~\beta_{3}~$and $\beta_{4}$ are constant, thus, in this
consideration coefficients $\beta_{1},~\beta_{2},~\beta_{3}~$and $\beta_{4}$
are promoted to functions of the scalar field $\phi\left(  x^{\mu}\right)  $.
This specific gravitational model is of special interest because in the case
of a Friedmann--Lema\^{\i}tre--Robertson--Walker (FLRW) universe provides two
periods of inflation, the slow-roll epoch and a second inflationary era which
follows from the domination of the Lorentz violating terms. Exact and analytic
solutions for this gravitational theory were found in \cite{ea1} for the
homogeneous and isotropic FLRW universe and in \cite{ea2} for the isotropic
and inhomogeneous Bianchi I spacetime.

In this work, we extend the analysis of the previous works by investigating
the existence of exact solutions when the underlying space is then
Kantowski-Sachs spacetime with the line-element:
\begin{equation}
ds^{2}=-N\left(  t\right)  dt^{2}+e^{2\lambda\left(  t\right)  }\left(
e^{2\beta\left(  t\right)  }dx^{2}+e^{-\beta\left(  t\right)  }\left(
d\theta^{2}+\sin^{2}\theta~d\varphi^{2}\right)  \right)  . \label{ac.03}%
\end{equation}
Function $N\left(  t\right)  $ is the lapse function, $e^{\lambda\left(
t\right)  }$ is the radius of the three dimensional space and $\beta\left(
t\right)  $ is the anisotropic parameter.

In the previous studies \cite{ea1,ea2} it is demonstrated that the
gravitational field equations for the Action Integral (\ref{ac.01}) can be
reproduced by the variation of a point-like Lagrangian with respect to
dynamical variables which are the unknown functions of the spacetime, that is,
$N\left(  t\right)  ,~\lambda\left(  t\right)  ,~\beta\left(  t\right)  $ and
the scalar field $\phi\left(  t\right)  $. At this point, we remark that the
scalar field $\phi$ is assumed to inherit all the isometries of the
Kantowski-Sachs spacetime.

By identifying the aether field with the velocity of a comoving observer, that
is $u_{\mu}=N\delta_{t}^{\mu},~$the point-like Lagrangian which produces the
gravitational field equations is
\begin{equation}
L\left(  N,\lambda,\dot{\lambda},\beta,\dot{\beta},\phi,\dot{\phi}\right)
=\frac{e^{3\lambda}}{N}\left(  -3F\left(  \phi\right)  \dot{\lambda}^{2}%
+\frac{3}{8}M\left(  \phi\right)  \dot{\beta}^{2}+\frac{1}{2}\dot{\phi}%
^{2}\right)  -Ne^{3\lambda}\left(  V\left(  \phi\right)  -e^{\beta-2\lambda
}\right)  , \label{ac.04}%
\end{equation}
where functions~$F\left(  \phi\right)  ,~M\left(  \phi\right)  $ are related
with the coefficient functions $\beta,$ as follows
\begin{equation}
F\left(  \phi\right)  =\left(  1+\beta_{1}\left(  \phi\right)  +3\beta
_{2}\left(  \phi\right)  +\beta_{3}\left(  \beta\right)  \right)  ,
\label{ac.05}%
\end{equation}%
\begin{equation}
M\left(  \phi\right)  =2\left(  1-2\left(  \beta_{1}\left(  \phi\right)
+\beta_{3}\left(  \phi\right)  \right)  \right)  . \label{ac.06}%
\end{equation}

The gravitational field equations are equivalent to
\begin{equation}
F\left(  \phi\right)  \left(  2\ddot{\lambda}+3\dot{\lambda}^{2}\right)
+\frac{3}{8}M\left(  \phi\right)  \dot{\beta}^{2} + \frac{1}{2}\dot{\phi}%
^{2}-V\left(  \phi\right)  +2F\left(  \phi\right)  _{,\phi}\dot{\lambda}%
\dot{\phi} +\frac{1}{3}e^{\beta-2\lambda}=0, \label{ac.08}%
\end{equation}%
\begin{equation}
M\left(  \phi\right)  \ddot{\beta}+3M\left(  \phi\right)  \dot{\lambda}%
\dot{\beta}+M\left(  \phi\right)  _{,\phi}\dot{\beta}\dot{\phi}-\frac{4}%
{3}e^{\beta-2\lambda}=0, \label{ac.09}%
\end{equation}%
\begin{equation}
\ddot{\phi}+3\dot{\lambda}\dot{\phi}+V\left(  \phi\right)  _{,\phi}+3F\left(
\phi\right)  _{,\phi}\dot{\lambda}^{2}-\frac{3}{8}M\left(  \phi\right)
_{,\phi}\dot{\beta}^{2}=0, \label{ac.10}%
\end{equation}
with constraint equation
\begin{equation}
-3F\left(  \phi\right)  \dot{\lambda}^{2}+\frac{3}{8}M\left(  \phi\right)
\dot{\beta}^{2}+\frac{1}{2}\dot{\phi}^{2}+V\left(  \phi\right)  -e^{\beta
-2\lambda}=0, \label{ac.07}%
\end{equation}
where the lapse function $N\left(  t\right)  =1$ is selected.

The field equations form a three-dimensional system with three unknown
functions, namely the functions $F\left(  \phi\right)  ,~M\left(  \phi\right)
$ and $V\left(  \phi\right)  $, and one conservation law, the constraint
equation (\ref{ac.07}). The dynamical system is nonlinear and a selection rule
should be applied in order to specify the unknown functions and construct
exact solutions.

Specific functional forms of $F\left(  \phi\right)  ,~M\left(  \phi\right)  $
and $V\left(  \phi\right)  $ are investigated such that the gravitational
field equations to admit additional conservation laws which can lead to
Liouville integrable models. The latter dynamical systems admit solutions
which can be expressed with the use of closed-form functions, that is, exact
solutions, or with the use of algebraic conditions. This approach is widely
applied in various alternative theories of gravity with interesting results.
It is also the method which is utilized in \cite{ea1,ea2} for the
determination of exact and analytic solutions. Another interesting
characteristic of this selection rule is of geometric origin, because there is
a one to one relation between the conservation laws and the geometry of the
minisuperspace which defines the kinetic part of the point-like Lagrangian
(\ref{ac.04}) for more details the reader is referred to the discussion in
\cite{sbns}.

\section{Exact solutions}

In this section the conservation laws of the field equations for specific
forms of the unknown functions are determined. Subsequently, they are applied
to derive exact solutions. In order to infer about the Liouville integrability
of the dynamical system there are needed at least two additional conservation
laws to be determined, which are independent and in involution with the
constraint equation (\ref{ac.07}).

The unknown functions $F\left(  \phi\right)  ,~M\left(  \phi\right)
$  and $V\left(  \phi\right)  $  are constrained by the
requirement of the existence of additional conservation laws. This requirement
 is equivalent to the existence of a symmetry vector field for the field
equations, which means that the specific functional forms $F\left(
\phi\right)  ,~M\left(  \phi\right)  $  and  $V\left(  \phi\right)
$ which are studied below are determined with the use of the symmetry
conditions and are not defined a priori by hand. In particular, we perform a
classification of the symmetries of the gravitational field equations as
defined by Ovsiannikov \cite{ovv}.

\subsection{Case A: $F\left(  \phi\right)  =M\left(  \phi\right)  $}

When $F\left(  \phi\right)  =M\left(  \phi\right)  $, the only unknown
functions are the $F\left(  \phi\right)  $ and the scalar field potential
$V\left(  \phi\right)  .$ We follow the procedure described in \cite{mtssym}
to determine conservation laws for the field equations (\ref{ac.08}%
)-(\ref{ac.07}).

Thus for $F\left(  \phi\right)  =\phi^{2}$ and $V\left(  \phi\right)  =V_{0}$,
the field equations admit the additional conservation law%
\begin{equation}
I_{1}=e^{3\lambda}\phi\left(  \phi\left(  \dot{\beta}-4\dot{\lambda}\right)
-2\dot{\phi}\right)  \label{ac.11}%
\end{equation}
generated by the Noether point symmetry $X_{1}=-\frac{3}{2} t\partial
_{t}+\frac{1}{2}\partial_{\lambda}+\partial_{\beta}-\frac{3}{2}\phi
\partial_{\phi}$. While the integrability of this model cannot be inferred,
the Noether symmetry can be used to determine an exact solutions. Indeed by
using the Lie invariants of the Lie symmetry vector it follows the exact
solutions%
\begin{equation}
\lambda\left(  t\right)  =-\ln t,~\beta\left(  t\right)  =-2\ln t,~\phi\left(
t\right)  =\frac{t}{\sqrt{3}}~\text{\ with }V_{0}=\frac{4}{3}\text{.}
\label{ac.12}%
\end{equation}

This is an anisotropic solution where the line element (\ref{ac.03}) is
written%
\begin{equation}
ds^{2}=-dt^{2}+t^{-6}dx^{2}+d\theta^{2}+\sin^{2}\theta d\varphi^{2}.
\label{ac.13}%
\end{equation}

However, in the special limit where $V\left(  \phi\right)  =0$, the analysis
differs. Specifically, in the case of the massless scalar field the
gravitational field equations admit the conservation laws%
\begin{equation}
\bar{I}_{1}=\phi^{2}e^{3\lambda}\left(  2\dot{\lambda}+\dot{\beta}\right)  ,
\label{ac.14}%
\end{equation}%
\begin{equation}
I_{2}=\phi e^{3\lambda}\left(  3\phi\dot{\lambda}+\dot{\phi}\right)  ,
\label{ac.15}%
\end{equation}
and%
\begin{equation}
I_{3}=\phi e^{3\lambda}\left(  \phi\left(  3\beta-2\ln\phi\right)
\dot{\lambda}-\phi\dot{\beta}\left(  3\lambda+\ln\phi\right)  +\dot{\phi
}\left(  \beta+2\lambda\right)  \right)  . \label{ac.17}%
\end{equation}
Remarkably, the conservation laws $\bar{I}_{1},~I_{2}$ are in involution, that
is, $\left\{  \bar{I}_{1},I_{2}\right\}  =0$, where $\left\{  ,\right\}  $ is
the Poisson bracket. Hence, it can be inferred that the gravitational field
equations form an integrable dynamical system.

By applying the change of variables%
\[
\beta=-4\lambda+u,~\phi=e^{v-2\lambda},
\]
the point-like Lagrangian (\ref{ac.04}) for $N\left(  t\right)  =\phi
^{2}e^{3\lambda}$ simplifies to%
\begin{equation}
L\left(  \lambda,\dot{\lambda},u,\dot{u},v,\dot{v}\right)  =5\dot{\lambda}%
^{2}+\frac{3}{8}\dot{u}^{2}+\frac{1}{2}\dot{v}^{2}-3\dot{u}\dot{\lambda}%
-2\dot{v}\dot{\lambda}+e^{u+2v-4\lambda}%
\end{equation}
from where we can write the Hamiltonian function%

\begin{equation}
\mathcal{H}\equiv-\frac{1}{12}\left(  8p_{u}^{2}-2p_{v}^{2}+4p_{v}p_{\lambda
}+p_{\lambda}^{2}+8p_{u}\left(  2p_{v}+p_{\lambda}\right)  \right)
-e^{u+2v-4\lambda}=0
\end{equation}
where~$p_{\lambda}=10 \dot{\lambda}-3\dot{u}-2\dot{v},~p_{u}=\frac{3}{4}%
\dot{u}-3 \dot{\lambda}$ and $p_{v}= \dot{v}-2 \dot{\lambda}$.

The conservation laws $\bar{I}_{1},~I_{2}$ become (note that due to changing
the time gauge, the corresponding expressions in the right hand side of
\eqref{ac.14}, \eqref{ac.15} linear in the velocity need to be multiplied by
$N^{-1}$):
\begin{equation}
\label{barI1}\bar{I}_{1}=-\frac{1}{3} \left(  2p_{v}+ p_{\lambda} \right)  ,
\end{equation}%
\begin{equation}
\label{I2}I_{2}=-\frac{1}{2}\left(  4p_{u}+p_{\lambda}\right)  .
\end{equation}

Therefore, with the use of the latter conservation laws we can solve the
Hamilton-Jacobi equation and reduce the order of the dynamical system and
write the analytic solution of the problem. In the simplest case where
$\bar{I}_{1}=0~,~I_{2}=0$, the action $S\left(  \lambda,u,v\right)  $ is%
\begin{equation}
S\left(  \lambda,u,v\right)  =\sqrt{3} e^{\frac{u}{2}+v-2\lambda}%
\end{equation}
which gives the reduced system%
\begin{equation}
\label{firstordeqs}\dot{\lambda}=-\frac{1}{\sqrt{3}}e^{\frac{u}{2}+v-2\lambda
}~,~\dot{u}=-\frac{2}{\sqrt{3}}e^{\frac{u}{2}+v-2\lambda}~,~\dot{v}=\frac
{1}{\sqrt{3}} e^{\frac{u}{2}+v-2\lambda}.
\end{equation}

Thus we can write the solution in terms of the radius $\lambda$, that is%
\begin{equation}
\frac{du}{d\lambda}=2~,~\frac{dv}{d\lambda}=-1
\end{equation}
from where we find $u\left(  \lambda\right)  =2\lambda+u_{0}~,~v=-\lambda
+v_{0}$, hence, $\beta=-2\lambda+u_{0}~,~\phi=e^{-3\lambda+v_{0}}$, and
$N\left(  \lambda\right)  =e^{-3\lambda+2v_{0}}$. Finally, from the first of
\eqref{firstordeqs} we are able to deduce $\lambda$ which reads
\begin{equation}
\lambda(t) = \frac{1}{2} \ln\left(  \pm\frac{2 e^{\frac{u_{0}}{2}+v_{0}}%
}{\sqrt{3}} t+ \lambda_{0} \right)  .
\end{equation}

We can transform the solution so as for the metric to be expressed with
respect to the cosmic time $\tau$ for which $N(\tau)=1$. With the help of a
transformation $t\mapsto\tau$ with
\begin{equation}
t= \pm\frac{\sqrt{3}}{2} e^{-\frac{3 u_{0}}{2}-v_{0}} \left(  \frac{3 e^{2
v_{0}}}{\tau^{2}}- \lambda_{0} e^{u_{0}} \right)
\end{equation}
the final solution reads
\begin{equation}
ds^{2} = -d\tau^{2} + \tau^{2} dx^{2} + \frac{e^{\alpha}}{\tau^{4}} \left(
d\theta^{2}+ \sin^{2}\theta d\varphi^{2} \right)  ,
\end{equation}
where also an appropriate scaling in the $x$ variable and a reparametrization
of the constants $9 e^{4 v_{0}-3 u_{0}}= e^{\alpha}$ has taken place in order
to simplify the line element. The corresponding massless scalar field is given
in this time gauge by $\phi(\tau) = \frac{e^{-\frac{\alpha}{2}} }{\sqrt{3}%
}\tau^{3}$.

In the most general case where $\bar{I}_{1}~I_{2}\neq0$ the solution of the
Hamilton- Jacobi equation is expressed as follows
\begin{equation}%
\begin{split}
S\left(  \lambda,u,v\right)  =  &  \frac{1}{4}\Bigg[\frac{3}{2}\left(
u-2(2\lambda+v)\right)  \bar{I}_{1}-(3u+2v-4\lambda)I_{2}+\sqrt
{48e^{u+2v-4\lambda}-S_{1}(\bar{I}_{1},I_{2})}\\
&  -\sqrt{S_{1}(\bar{I}_{1},I_{2})}\arctan\left(  \frac{\sqrt
{48e^{u+2v-4\lambda}-S_{1}(\bar{I}_{1},I_{2})}}{\sqrt{S_{1}(\bar{I}_{1}%
,I_{2})}}\right)  \Bigg]
\end{split}
\end{equation}
where $S_{1}\left(  \bar{I}_{1},I_{2}\right)  =3\left(  3\bar{I}_{1}%
^{2}-12\bar{I}_{1}I_{2}-4I_{2}^{2}\right)  $. Although deriving the solution
in a similar manner becomes more cumbersome, we can use the third integral of
motion \eqref{ac.17}, which in phase space variables is written as
\begin{equation}
I_{3}=\frac{1}{6}\left(  (8\lambda-3u+2v)p_{\lambda}-12(u-2\lambda
)p_{u}+4(\lambda+v)p_{v}\right)  , \label{I3b1}%
\end{equation}
to obtain additional information. If we exploit \eqref{barI1} and \eqref{I2}
with the means to substitute two of the momenta in expression \eqref{I3b1}, we
observe that the latter leads to an algebraic relation among the configuration
space variables. In particular we get
\begin{equation}
u=\frac{(\bar{I}_{1}+2I_{2})\lambda+\bar{I}_{1}v+I_{3}}{I_{2}}, \label{solu}%
\end{equation}
for which we assume from now on that $I_{2}\neq0$. If we turn back to the
expression \eqref{I2} for the integral of motion $I_{2}$ we see that, in the
velocity phase space, it is written as
\begin{equation}
I_{2}\dot{\lambda}+\dot{v}\Rightarrow v(t)=I_{2}t-\lambda+v_{0}. \label{solv}%
\end{equation}
At this point we need to determine $\lambda$. We have no additional integrals
of motion to exploit so we turn to the field equations. Under the conditions
we have imposed on $F$, $M$, $V$ - together with the relations obtained
\eqref{solu}, \eqref{solv} - equation \eqref{ac.10} becomes (remember that it
is needed to reinstate the lapse function $N$ in \eqref{ac.10} and then apply
the current gauge fixing condition $N=\phi^{2}e^{3\lambda}$):
\begin{equation}
12\ddot{\lambda}-12(\bar{I}_{1}+2I_{2})\dot{\lambda}+24\dot{\lambda}^{2}%
+3\bar{I}_{1}^{2}+4I_{2}^{2}=0
\end{equation}
which is essentially a first order relation due to not involving $\lambda$
itself. It can be easily integrated to yield
\begin{equation}
\lambda(t)=\lambda_{1}+\frac{(\bar{I}_{1}+2I_{2})}{4}t+\frac{1}{2}\ln\left[
\cos\left(  \frac{1}{6}\sqrt{S_{1}(\bar{I}_{1},I_{2})}(t-\lambda_{2})\right)
\right]  ,
\end{equation}
where $\lambda_{1},\lambda_{2}$ are constants of integration. Finally, the
constraint relation \eqref{ac.07} (again the new lapse has to be taken into
count) sets a condition among the constants of integration
\begin{equation}
\lambda_{1}=\frac{1}{4}\ln\left[  \frac{48e^{\frac{\bar{I}_{1}v_{0}%
+2I_{2}v_{0}+I_{3}}{I_{2}}}}{S_{1}(\bar{I}_{1},I_{2})}\right]  .
\end{equation}
With the help of the above, we can write the corresponding line element as
\begin{equation}
ds^{2}=-\frac{e^{\alpha t}}{\left[  \cos\left(  \sqrt{\tilde{S}_{1}}t\right)
\right]  ^{3}}dt^{2}+\frac{e^{-\alpha t}}{\cos\left(  \sqrt{\tilde{S}_{1}%
}t\right)  }dx^{2}+e^{\frac{1}{2}\left(  \alpha+\sqrt{3}\sqrt{\alpha
^{2}-6\tilde{S}_{1}}\right)  t}\left[  \beta_{0}\cos\left(  \sqrt{\tilde
{S}_{1}}t\right)  \right]  ^{2}\left(  d\theta^{2}+\sin^{2}\theta d\varphi
^{2}\right)  ,
\end{equation}
where, for simplification, again we performed a scaling in $x$ together with
an appropriate transformation in time $t\mapsto e^{\frac{\beta_{0}}{2}%
}t+\lambda_{2}$ and a reparametrization of constants $(\bar{I}_{1},I_{2}%
,I_{3},\lambda_{2})\mapsto(\tilde{S}_{1},\alpha,\beta_{1},\beta_{0})$ as
\begin{equation}%
\begin{split}
&  \tilde{S}_{1}=\frac{e^{\beta_{1}}S_{1}(\bar{I}_{1},I_{2})}{36},\quad
\alpha=-\frac{1}{2}e^{\frac{\beta_{1}}{2}}(3\bar{I}_{1}-2I_{2}),\quad
I_{3}=\frac{2}{3}I_{2}\left[  \frac{\beta_{1}}{4}+\alpha e^{-\frac{\beta_{1}%
}{2}}\left(  \frac{v_{0}}{I_{2}}+\lambda_{2}\right)  +\ln\left(  \frac
{3\sqrt{3}}{8}e^{-\frac{3\beta_{1}}{4}}\tilde{S}_{1}^{\frac{3}{2}}\right)
\right] \\
&  \lambda_{2}=\frac{4\left[  \ln\left(  \frac{3\beta^{2}\tilde{S}_{1}}%
{4}\right)  -\beta_{1}-2v_{0}\right]  }{e^{-\frac{\beta}{2}}\left(
2\alpha+\sqrt{12}\sqrt{\alpha^{2}-6\tilde{S}_{1}}\right)  }.
\end{split}
\label{constantspar}%
\end{equation}
Under these changes the scalar field is given by
\begin{equation}
\phi(t)=\frac{2e^{\frac{1}{4}\left(  \alpha-\sqrt{3}\sqrt{\alpha^{2}%
-6\tilde{S}_{1}}\right)  t}}{\sqrt{3}\beta_{0}\sqrt{\tilde{S}_{1}}\left[
\cos\left(  \sqrt{\tilde{S}_{1}}t\right)  \right]  ^{\frac{3}{2}}}.
\end{equation}
Even though the above solution was extracted under the assumption $I_{2}\neq
0$, if we enforce from the beginning $I_{2}=0$ and follow a similar procedure,
we are led to the same form for the line element but by the application of a
different reparametrization for the constants of integration.

\subsection{Case B: $F\left(  \phi\right)  \neq M\left(  \phi\right)  $}

We continue our analysis by assuming $F\left(  \phi\right)  \neq M\left(
\phi\right)  $. By applying the algorithm described in \cite{mtssym} we find
that the gravitational field equations admit additional conservation laws
linear in the momentum when $V\left(  \phi\right)  =0$, and Subcase B1 with
$\left\{  F\left(  \phi\right)  ,M\left(  \phi\right)  \right\}  =\left\{
F_{0},e^{\frac{2}{3}\sqrt{\frac{6}{F_{0}}}\phi}\right\}  $ or Subcase B2 where
$\left\{  F\left(  \phi\right)  ,M\left(  \phi\right)  \right\}  =\left\{
\cos\left(  \frac{2}{3}\sqrt{\frac{3}{F_{0}}}\phi\right)  +1,M_{0}\right\}  $,
or Subcase $B_{3}$ with $\left\{  F\left(  \phi\right)  ,M\left(  \phi\right)
\right\}  $ arbitrary.

\subsubsection{Subcase B1}

In the first Subcase the gravitational field equations admit the additional
conservation laws%
\begin{equation}
\label{I4int}I_{4}=\frac{e^{3\lambda}}{N}\left(  2F_{0}\dot{\lambda}%
+e^{\frac{2}{3}\sqrt{\frac{6}{F_{0}}}\phi}\dot{\beta}\right)  ,
\end{equation}%
\begin{equation}
\label{I5int}I_{5}=\frac{e^{\lambda+\sqrt{\frac{6}{F_{0}}}\frac{\phi}{3}}}%
{N}\left(  \sqrt{6F_{0}}\dot{\lambda}+\dot{\phi}\right)  ,
\end{equation}
where we have expressed them in an arbitrary gauge $N=N(t)$.

The two new conservation laws are not in involution, hence we can not infer
about the integrability of the dynamical system. However, we are able to
integrate the equations in the special case where $I_{5}=0$. Let us choose to
work in the time gauge $N= e^{3\lambda+ \frac{2}{3} \sqrt{\frac{6}{F_{0}}}
\phi}$. Then, the integral of motion \eqref{I4int} leads to
\begin{equation}
\label{solb1phi}\phi(t) = \frac{1}{2} \sqrt{\frac{3F_{0}}{2}} \ln\left(
\frac{2 F_{0} \dot{\lambda}}{I_{4}- \dot{\beta}} \right)  .
\end{equation}
Use of the above expression into \eqref{I5int} yields an equation easily
integrated with respect to $\beta$ with solution
\begin{equation}
\label{solb1beta}\beta(t) = \int\left[  I_{4} -\frac{1}{12} e^{4 \lambda}
\left(  \beta_{1} +4 I_{5} t\right)  ^{2} \dot{\lambda} \right]  dt .
\end{equation}
With the help of \eqref{solb1phi}, \eqref{solb1beta} and introducing a
function $\omega(t)$ as $\lambda(t)= \frac{1}{4} \ln\omega$, the
Euler-Lagrange equation for $\phi$ becomes
\begin{equation}
\label{equomega}(\beta_{1}+4 I_{5} t)^{2} \ddot{\omega} + \frac{1}{96}
(\beta_{1}+4 I_{5} t)^{4} \dot{\omega}^{2} -(\beta_{1}+4 I_{5} t) \left(
\beta_{1} I_{4}+4 I_{5} (I_{4} t-4)\right)  \dot{\omega} + 32 I_{5}^{2}
\omega+ 24 I_{4}^{2} =0.
\end{equation}
In the special case where $I_{5}=0$ the above equation has the simple
solution
\begin{equation}
\omega(t) = \frac{48}{\beta_{1}^{2}} \left(  I_{4} t+2 \ln\left(  t-\beta
_{1}^{2} \omega_{1}\right)  \right)  + \omega_{2} .
\end{equation}
Of course, solving \eqref{equomega} is not enough, we need to make sure that
the constraint equation is also satisfied. The latter leads to the additional
condition among constants $\beta_{1} = 4 e^{\frac{\beta_{2}}{2}}\sqrt{F_{0}}$.

After a scaling in $x$, a transformation
\begin{equation}
t \mapsto\frac{2^{3/4} e^{\frac{3 \beta_{2}}{4}} }{3 F_{0}^{1/4}} t +
\omega_{1}%
\end{equation}
and reparametrizations of the constants of integration $(\omega_{1}%
,I_{4})\mapsto(t_{0},\tilde{I}_{4})$:
\begin{equation}
\omega_{1} = \frac{1}{6I_{4}} \left(  t_{0} -9 \beta_{2}-2 e^{\beta_{2}} F_{0}
\omega_{2} + 3 \ln\left(  \frac{81 F_{0}}{8}\right)  \right)  , \quad I_{4} =
\frac{e^{-\frac{3 \beta_{2}}{4}} \tilde{I}_{4} F_{0}^{1/4}}{2\times2^{3/4}},
\end{equation}
the final solution is expressed as
\begin{equation}
ds^{2} = - \frac{1}{\sqrt{12 \ln(t) +\tilde{I}_{4} t + t_{0}}} dt^{2} +
\frac{\sqrt{12 \ln(t) +\tilde{I}_{4} t + t_{0}}}{t^{4}} dx^{2} + \frac{2
t^{2}}{9 F_{0}}\sqrt{12 \ln(t) +\tilde{I}_{4} t + t_{0}} \left(  d\theta^{2} +
\sin^{2}\theta d\varphi^{2} \right)  ,
\end{equation}
corresponding to the scalar field
\begin{equation}
\phi(t) = \frac{1}{2} \sqrt{\frac{3F_{0}}{2}} \ln\left(  \frac{3 F_{0}}{12
\ln(t) + \tilde{I}_{4} t +t_{0}} \right)  .
\end{equation}

As long as the generic case where $I_{5}\neq0$ is concerned, we may just
notice, that there exists a transformation that can render \eqref{equomega}
autonomous or, alternatively, reduce its order and replace it with an Abel
equation. In the first case application of the transformation $(t,\omega
(t))\mapsto(s,\zeta(s))$
\begin{equation}
\label{pointtr1}t = \frac{e^{4 I_{5} s }-\beta_{1}}{4 I_{5}} , \quad\omega=
\frac{e^{-8 I_{5} s } \left(  I_{5} \zeta-12 I_{4} \left(  e^{4 I_{5} s
}-\beta_{1}\right)  \right)  }{I_{5}}%
\end{equation}
leads to
\begin{equation}
\zeta^{\prime\prime}+ \frac{1}{96} \zeta^{\prime2 }-\left(  2 \beta_{1}
I_{4}+\frac{1}{6} I_{5} \zeta+4 I_{5}\right)  \zeta^{\prime}+ \frac{2
I_{5}^{2}}{3} \zeta^{2} + 16 \beta_{1} I_{4} I_{5} \zeta+ 96 \beta_{1}^{2}
I_{4}^{2} =0,
\end{equation}
where the prime denotes now the derivatives with respect to the new variable
$s$. The obvious solution $\zeta=-\frac{12 \beta_{1} I_{4}}{I_{5}}$ does not
lead to a valid result since the constraint then demands $I_{4}=0$, which
makes the corresponding $\lambda(t) = \frac{1}{3} \ln\left[  \frac{-12 I_{4}%
}{I_{5}(4 I_{5}t+\beta_{1})}\right]  $ diverge.

On the other hand, by introducing the transformation (interchanging $\zeta$
and $s$ in \eqref{pointtr1})
\begin{equation}
t = \frac{e^{4 I_{5} \zeta}-\beta_{1}}{4 I_{5}} , \quad\omega= \frac{e^{-8
I_{5} \zeta} \left(  I_{5} s -12 I_{4} \left(  e^{4 I_{5} \zeta}-\beta
_{1}\right)  \right)  }{I_{5}},
\end{equation}
together with the additional use of $\zeta=\int\! \chi(s) ds$, we obtain
\begin{equation}
\chi^{\prime}-\frac{1}{3} 2 (12 \beta_{1} I_{4}+I_{5} s )^{2} \chi^{3}
+\left(  2 \beta_{1} I_{4}+\frac{1}{6} I_{5} (s +24)\right)  \chi^{2} -
\frac{\chi}{96} =0.
\end{equation}
Once more the obvious solution $\chi=0$ does not lead to a valid result since
it implies $\zeta=$const.$\Rightarrow t=$const.

\subsubsection{Subcase B2}

In the second case the additional conservation laws are derived to be
\begin{equation}
I_{6}=e^{3\lambda}\left(  2\left(  \cos\left(  \frac{2}{3}\sqrt{\frac{3}%
{M_{0}}}\phi\right)  +1\right)  \dot{\lambda}+M_{0}\dot{\beta}\right)
\end{equation}

\begin{equation}
I_{7}=e^{-\frac{\beta}{2}+3\lambda}\left(  3\sin\left(  \frac{1}{3}\sqrt
{\frac{3}{M_{0}}}\phi\right)  M_{0}\dot{\beta}+2\sqrt{3M_{0}}\cos\left(
\frac{1}{3}\sqrt{\frac{3}{M_{0}}}\phi\right)  \dot{\phi}\right)
\end{equation}
while the conservation laws are not in involution. Therefore, the
integrability of the field equations can not be further discussed.

\subsubsection{Subcase B3}

In the most general Subcase with two arbitrary functions $\left\{  F\left(
\phi\right)  ,M\left(  \phi\right)  \right\}  $ and massless scalar field,
that is, $V\left(  \phi\right)  =0$, the gravitational field equations admit
the additional conservation law%
\begin{equation}
I_{8}=e^{3\lambda}\left(  2F\left(  \phi\right)  \dot{\lambda}+M\left(
\phi\right)  \dot{\beta}\right)
\end{equation}
which includes the conservation laws $\bar{I}_{1},~I_{4}$ and $I_{6}$.

In the following analysis the asymptotic behaviour of the equation's solutions
is studied, and a detailed study of the stationary points for the field
equations is performed.

\section{Asymptotic behaviour}

In order to study the evolution of the dynamics of the gravitational model the
following dimensionless variables are defined:%
\begin{equation}
x=\frac{\dot{\phi}}{\sqrt{6F}\dot{\lambda}},~y=\sqrt{\frac{V}{3F\dot{\lambda
}^{2}}},~\Sigma=\frac{1}{2}\sqrt{\frac{M}{2F}}\frac{\dot{\beta}}{\dot{\lambda
}},~R^{\left(  3\right)  }=\frac{e^{\beta-2\lambda}}{3F\dot{\lambda}^{2}}
\label{dyn.01}%
\end{equation}

The model A, with $M\left(  \phi\right)  =F\left(  \phi\right)  $ and
$F\left(  \phi\right)  =\phi^{2}$ is discussed. The gravitational field
equations are written in the form of the following algebraic-differential
system%
\begin{equation}
\frac{dx}{d\lambda}=-\sqrt{\frac{3}{2}}\left(  2+\mu y^{2}-2\Sigma^{2}\right)
+x\left(  x\left(  \sqrt{6}+2x\right)  -2-y^{2}+2\Sigma^{2}\right)  ,
\label{dyn.02}%
\end{equation}%
\begin{equation}
\frac{dy}{d\lambda}=\frac{1}{2}y\left(  2+4x^{2}-2y^{2}+\sqrt{6}\left(
2+\mu\right)  x+4\Sigma^{2}\right)  , \label{dyn.04}%
\end{equation}%
\begin{equation}
\frac{d\Sigma}{d\lambda}=y^{2}\left(  \sqrt{2}-\Sigma\right)  +\left(
\sqrt{2}+2\Sigma\right)  \left(  x^{2}+\Sigma^{2}-1\right)  , \label{dyn.03}%
\end{equation}
with algebraic equation
\begin{equation}
1-x^{2}-y^{2}-\Sigma^{2}+R^{\left(  3\right)  }=0. \label{dyn.05}%
\end{equation}
The additional equation
\begin{equation}
\frac{d R^{\left(  3\right)  }}{d \lambda}= R^{\left(  3\right)  }\left[  2
\left(  {\Sigma} \left(  2 {\Sigma}+\sqrt{2}\right)  +x \left(  2 x+\sqrt
{6}\right)  -y^{2}\right)  \right]  ,
\end{equation}
is derived, from which it follows the sign invariance of $R^{\left(  3\right)
}$. Under the assumption $F>0$, the phase space will be given by the exterior
and the surface of the hemisphere:
\begin{align}
\{(x, y, \Sigma)\in\mathbb{R}^{3}: x^{2}+y^{2}+\Sigma^{2}\geq1, y\geq0\}.
\end{align}

For the scalar field the potential $V\left(  \phi\right)  =V_{0}\phi^{\mu}$
was assumed. The special case $\mu=0$ corresponds to the cosmological constant
term, while $y=0$ corresponds to the massless scalar field. Each point
$P=\left(  x\left(  P\right)  ,y\left(  P\right)  ,\Sigma\left(  P\right)
\right)  $ at which the right hand side of (\ref{dyn.02})-(\ref{dyn.04})
vanishes is a stationary point of the dynamical system and describes an
specific epoch of the cosmological evolution.

The stationary points of the field equations are calculated:%
\[
P_{1}^{\pm}[x]=\left(  x,0,\pm\sqrt{1-x^{2}}\right)  ,
\]%
\[
~P_{2}=\left(  -\sqrt{\frac{3}{2}},0,-\frac{\sqrt{2}}{2}\right)  ,
\]%
\[
P_{3}=\left(  -\frac{\mu+2}{\sqrt{6}},\sqrt{\frac{ 2-\mu\left(  \mu+4\right)
}{6}},0\right)  ,~
\]%
\[
P_{4}=\left(  -\frac{\sqrt{\frac{2}{3}}\left(  1+2\mu\right)  }{2+\mu\left(
\mu-1\right)  },\sqrt{\frac{\left(  4-3\mu\right)  \left(  4+\mu\left(
\mu-8\right)  \right)  }{3\left(  2+\mu\left(  \mu-1\right)  \right)  ^{2}}%
},\frac{2-\mu\left(  \mu+2\right)  }{\sqrt{2}\left(  2+\mu\left(
\mu-1\right)  \right)  }\right)  .
\]

Points $P_{1}^{\pm\ }[x]$ correspond to the same dynamics as in the case of an
anisotropic Bianchi I spacetime where $R^{\left(  3\right)  }\left(
P_{1}^{\pm}[x]\right)  =0$ and only the kinetic part of the scalar field
contributes to the cosmological fluid. The points are real when $\left\vert
x\right\vert \leq1$ while in the limit $\left\vert x\right\vert =1$, the
dynamics at $P_{1}^{\left(  \pm\right)  }[x]$ become that of spatially flat
FLRW space. \ The eigenvalues of the linearized system around the stationary
points are:%
\[
e_{1}\left(  P_{1}^{+}[x]\right)  =0,~e_{2}\left(  P_{1}^{+}[x]\right)
=3+\sqrt{\frac{3}{2}}\left(  \mu+2\right)  x,~e_{3}\left(  P_{1}%
^{+}[x]\right)  =2\left(  2+\sqrt{6}x+\sqrt{2\left(  1-x^{2}\right)  }\right)
\]
and%
\[
e_{1}\left(  P_{1}^{-}[x]\right)  =0,~e_{2}\left(  P_{1}^{-}[x]\right)
=3+\sqrt{\frac{3}{2}}\left(  \mu+2\right)  x,~e_{3}\left(  P_{1}%
^{-}[x]\right)  =2\left(  2+\sqrt{6}x-\sqrt{2\left(  1-x^{2}\right)  }\right)
.
\]
The lines $P_{1}^{\left(  \pm\right)  }[x]$ are normally-hyperbolic invariant
sets. Indeed, the parametric curves can be expressed as:
\[
\mathbf{r}(x)=\left(  x,0,\pm\sqrt{1-x^{2}}\right)  .
\]
Its tangent vector evaluated at a given $x$ is:
\[
\mathbf{r}^{\prime}(x)=\left(  1,0,\mp\frac{x}{\sqrt{1-x^{2}}}\right)  ,
\]
is parallel to the eigenvector corresponding to the zero eigenvalue, given
by:
\[
\mathbf{v}(x)=\left(  \mp\frac{\sqrt{1-x^{2}}}{x},0,1\right)  .
\]
In this particular case, the stability can be studied considering only the
signs of the real parts of the non-zero eigenvalues. In this way it is
concluded that $P_{1}^{+}[x]$ is a sink for:
\[
\sqrt{6}-2<\mu\leq4-2\sqrt{3},\quad-1\leq x<-\frac{\sqrt{6}}{\mu+2},
\]
or
\[
\mu>4-2\sqrt{3},\quad-1\leq x<\frac{1}{4}\left(  -\sqrt{2}-\sqrt{6}\right)  .
\]
$P_{1}^{+}[x]$ is a source for:
\[
\mu\leq-2-\sqrt{6},\quad\frac{1}{4}\left(  -\sqrt{2}-\sqrt{6}\right)
<x<-\frac{\sqrt{6}}{\mu+2},
\]
or
\[
-2-\sqrt{6}<\mu<4-2\sqrt{3},\quad\frac{1}{4}\left(  -\sqrt{2}-\sqrt{6}\right)
<x\leq1,
\]
or
\[
\mu\geq4-2\sqrt{3},\quad-\frac{\sqrt{6}}{\mu+2}<x\leq1.
\]
$P_{1}^{+}[x]$ is a saddle for:
\[
\mu=\sqrt{6}-2,\quad-1<x<\frac{1}{4}\left(  -\sqrt{2}-\sqrt{6}\right)  ,
\]
or
\[
\mu<\sqrt{6}-2,\quad-1\leq x<\frac{1}{4}\left(  -\sqrt{2}-\sqrt{6}\right)  ,
\]
or
\[
\sqrt{6}-2<\mu<4-2\sqrt{3},\quad-\frac{\sqrt{6}}{\mu+2}<x<\frac{1}{4}\left(
-\sqrt{2}-\sqrt{6}\right)  ,
\]
or
\[
\mu<-2-\sqrt{6},\quad-\frac{\sqrt{6}}{\mu+2}<x\leq1,
\]
or
\[
\mu>4-2\sqrt{3},\quad\frac{1}{4}\left(  -\sqrt{2}-\sqrt{6}\right)
<x<-\frac{\sqrt{6}}{\mu+2}.
\]

On the other hand, $P_{1}^{-}[x]$ is a sink for:
\[
\sqrt{6}-2<\mu\leq4+2 \sqrt{3}, \quad-1\leq x<-\frac{\sqrt{6}}{\mu+2},
\]
or
\[
\mu>4+2 \sqrt{3}, \quad-1\leq x<\frac{1}{4} \left(  \sqrt{2}-\sqrt{6}\right)
;
\]
$P_{1}^{-}[x]$ is a source for:
\[
\mu\leq-2-\sqrt{6}, \quad\frac{1}{4} \left(  \sqrt{2}-\sqrt{6}\right)
<x<-\frac{\sqrt{6}}{\mu+2},
\]
or
\[
-2-\sqrt{6}<\mu<4+2 \sqrt{3}, \quad\frac{1}{4} \left(  \sqrt{2}-\sqrt
{6}\right)  <x\leq1,
\]
or
\[
\mu\geq4+2 \sqrt{3}, \quad-\frac{\sqrt{6}}{\mu+2}<x\leq1.
\]
Finally, $P_{1}^{-}[x]$ is a saddle for:
\[
\mu<-2, \quad x>-\frac{\sqrt{6}}{\mu+2},
\]
or
\[
\mu>-2, \quad x<-\frac{\sqrt{6}}{\mu+2},
\]
or
\[
\frac{1}{4} \left(  \sqrt{2}-\sqrt{6}\right)  <x\leq1.
\]

For point $P_{2}$ is calculated $R^{\left(  3\right)  }\left(  P_{2}\right)
=1$, and because $\Sigma\left(  P_{2}\right)  \neq0$, the exact solution
describes a Kantowski-Sachs spacetime where only the kinetic part of the
scalar field contributes in the exact solution. The following eigenvalues of
the linearized system around $P_{2}$ are derived:%
\[
e_{1}\left(  P_{2}\right)  =2,~e_{2}\left(  P_{2}\right)  =2-\frac{3}{2}\mu,
~e_{3}\left(  P_{2}\right)  =2,
\]
from where it is concluded that the stationary point is a saddle point.

Point $P_{3}$ has~$\Sigma\left(  P_{3}\right)  =0$, and $R^{\left(  3\right)
}\left(  P_{3}\right)  =0$, which means that the exact solution at the point
approaches that of a spatially flat FLRW universe. The point is real when
$\left\vert \mu+2\right\vert <\sqrt{6}$. The parameter for the equation of
state for the cosmological fluid is $w_{tot}\left(  P_{3}\right)  =\frac{1}%
{3}\left(  \mu^{2}-7\right)  $ from where it is inferred that the exact
solution describes an accelerating universe when $-\sqrt{6}<\mu<-2+\sqrt{6}$.
The eigenvalues of the linearized system around $P_{3}$ are
\[
e_{1}\left(  P_{3}\right)  =-2+\mu\left(  \mu+2\right)  ,~e_{2}\left(
P_{3}\right)  =\frac{1}{2}\left(  -2+\mu\left(  \mu+4\right)  \right)
,~e_{3}\left(  P_{3}\right)  =\frac{1}{2}\left(  -2+\mu\left(  \mu+4\right)
\right)
\]
from where we conclude that the stationary point is an attractor when
$-1-\sqrt{3}<\mu<-2+\sqrt{6}$.

Finally for $P_{4}$ it is derived $R^{\left(  3\right)  }\left(  P_{4}\right)
=-\frac{\left(  4+\mu\left(  \mu-8\right)  \right)  \left(  \mu\left(
\mu+2\right)  -2\right)  }{2\left(  \mu\left(  \mu-1\right)  +2\right)  ^{2}}%
$, where the exact solution is that of a Kantowski-Sachs space. The point is
real for $\mu<2\left(  2-\sqrt{3}\right)  $ and $\frac{4}{3}<\mu<2\left(
2+\sqrt{3}\right)  $ while when $\mu=2\left(  2\pm\sqrt{3}\right)  $ the
evolution of the scale factors simulates those of an anisotropic Bianchi I
space, or when $\mu=-1-\sqrt{3}$ where the exact solution follows the
behaviour of a spatially flat FLRW spacetime. The eigenvalues of the
linearized system around $P_{4}$ are calculated:%
\[
e_{1}\left(  P_{4}\right)  =-1+\frac{7\mu-2}{2+\mu\left(  \mu-1\right)  }~
\]%
\[
e_{\pm}\left(  P_{4}\right)  =-\frac{8+\mu\left(  \mu\left(  \mu-7\right)
\left(  \mu-2\right)  -20\right)  }{2\left(  2+\mu\left(  \mu-1\right)
\right)  ^{2}}\pm\frac{1}{2}\sqrt{\Delta}%
\]
with $\Delta\left(  \mu\right)  =\left(  4+\mu\left(  \mu-8\right)  \right)
\left(  36+\mu\left(  3\mu\left(  4\mu+3\right)  -64\right)  \right)  $, from
where it is concluded that for $\mu<-1-\sqrt{3}$ the stationary point is an
attractor and the exact solution is stable.

\subsection{Compactification procedure}

In order to find a compact phase space for $R^{\left(  3\right)  }\geq0$ is
used the equation
\begin{equation}
1=\frac{x^{2}}{1+R^{\left(  3\right)  }}+\frac{y^{2}}{1+R^{\left(  3\right)
}}+\frac{\Sigma^{2}}{1+R^{\left(  3\right)  }}, \quad Z=\frac{1}%
{\sqrt{1+R^{\left(  3\right)  }}}, \label{dyn.38}%
\end{equation}
to define bounded variables. Indeed, choosing
\begin{equation}
\label{EQQ39}X= \frac{x}{\sqrt{1+ R^{\left(  3\right)  }}}, \quad Y= \frac{y
}{\sqrt{1+ R^{\left(  3\right)  }}}, \quad Z =\frac{1}{\sqrt{1+R^{\left(
3\right)  }}},
\end{equation}
with inverse functions
\begin{equation}
x= \frac{X}{Z}, \quad y= \frac{Y}{Z}, \quad\Sigma=\frac{\sqrt{1-X^{2}-Y^{2}}%
}{Z}, \quad R^{\left(  3\right)  }= \frac{1-Z^{2}}{Z^{2}},
\end{equation}
the following dynamical system is obtained:
\begin{align}
&  Z \frac{d X}{d \lambda}= X \left(  \sqrt{2} \left(  Z^{2}-1\right)
\sqrt{1-X^{2}-Y^{2}}-3 Y^{2} Z\right)  +\sqrt{6} X^{2} \left(  Z^{2}-1\right)
-\sqrt{\frac{3}{2}} \left(  (\mu+2) Y^{2}+2 \left(  Z^{2}-1\right)  \right)
,\label{eq42}\\
&  Z \frac{d Y}{d \lambda}= \frac{1}{2} Y \left(  2 \sqrt{2} Z^{2}
\sqrt{1-X^{2}-Y^{2}}-2 \sqrt{2} \sqrt{1-X^{2}-Y^{2}}+\sqrt{6} X \left(  \mu+2
Z^{2}\right)  -6 \left(  Y^{2}-1\right)  Z\right)  ,\label{eq43}\\
&  Z \frac{d Z}{d \lambda}= -\left(  1-Z^{2}\right)  \left(  \sqrt{2} Z
\sqrt{1-X^{2}-Y^{2}}+\sqrt{6} X Z-3 Y^{2}+2\right)  , \label{eq44}%
\end{align}
defined on the compact phase space
\begin{align}
\{(X,Y,Z)\in\mathbb{R}^{3}: X^{2}+Y^{2}\leq1, \quad Y\geq0, \quad0\leq
Z\leq1\}.
\end{align}

The isotropic universes corresponds to the invariant circle $X^{2}+Y^{2}=1$.
Using the parametrization
\begin{equation}
X= \cos\Phi, \quad Y= \sin\Phi, \quad\Phi\in[0,\pi],
\end{equation}
the dynamics over the invariant circle is given by
\begin{align}
&  \frac{d \Phi}{d \lambda}= \frac{1}{2 Z} \sin(\Phi) \left(  \sqrt{6} \left(
\mu+2 Z^{2}\right)  +6 Z \cos(\Phi)\right)  ,\label{eq47}\\
&  \frac{d Z}{d \lambda}= -\frac{1}{2 Z} \left(  1-Z^{2}\right)  \left(  3
\cos(2 \Phi)+2 \sqrt{6} Z \cos(\Phi)+1\right)  . \label{eq48}%
\end{align}
The isotropic universes are calculated to be: \newline%
\[
P_{1+}=P_{1}^{\left(  +\right)  }[1]=P_{1}^{\left(  -\right)  }[1]: (\Phi, Z)=
\left(  2 n \pi, 1\right)  , \{x=1, y=0,\Sigma= 0,R^{(3)}=0\},
\]
\[
P_{1-}=P_{1}^{\left(  +\right)  }[-1]=P_{1}^{\left(  -\right)  }[-1]: (\Phi,
Z)=\left(  (2 n+1) \pi, 1\right)  , \{x=-1,y= 0,\Sigma= 0,R^{(3)}= 0\},
\]
\[
P_{3}: (\Phi, Z)= \left(  \tan^{-1}\left(  -\frac{\mu+2}{\sqrt{6}},\frac
{\sqrt{2-\mu(\mu+4)}}{\sqrt{6}}\right)  +2 n \pi,1\right)  , \left\{  x=
-\frac{\mu+2}{\sqrt{6}},y= \frac{\sqrt{2-\mu(\mu+4)}}{\sqrt{6}},\Sigma=
0,R^{(3)}=0\right\}  ,
\]
where $n \in\mathbb{Z}$ and $\tan^{-1}[x,y]$ gives the arc tangent of $y/x$,
taking into account on which quadrant the point $(x,y)$ is in. When
$x^{2}+y^{2}=1$, $\tan^{-1}[x,y]$ gives the number $\Phi$ such as $x=\cos\Phi$
and $y=\sin\Phi$. \newline%
\[
P_{5}: (\Phi, Z)= \left(  (2 n+1) \pi,\sqrt{\frac{2}{3}}\right)  , \left\{  x=
-\sqrt{\frac{3}{2}}, y=0,\Sigma=0, R=\frac{1}{2}\right\}  .
\]
$P_{5}$ is not a stationary point of the original system (\ref{dyn.02}%
)-(\ref{dyn.04}) since $\Sigma^{\prime}(P_{5})=\frac{\sqrt{2}}{2}$, but it is
a stationary point of \eqref{eq42}, \eqref{eq43}, \eqref{eq44}.\newline%
\[
P_{6}: (\Phi, Z)= \left(  \tan^{-1}\left(  \frac{1}{\sqrt{3} \sqrt{1-\mu}%
},\frac{\sqrt{2-3 \mu}}{\sqrt{3} \sqrt{1-\mu}}\right)  + 2 n \pi,-\frac{\mu
}{\sqrt{2(1- \mu) }}\right)  ,
\]
\[
\left\{  x= -\frac{\sqrt{\frac{2}{3}}}{\mu},y= -\frac{\sqrt{\frac{4}{3}-2 \mu
}}{\mu},\Sigma= 0, R^{(3)}= \frac{2(1-\mu) }{\mu^{2}}-1\right\}  , -1-\sqrt
{3}\leq\mu<0.
\]
$P_{6}$ is not a stationary point of the original system (\ref{dyn.02}%
)-(\ref{dyn.04}) since $\Sigma^{\prime}(P_{6})=\frac{\sqrt{2} (2-\mu(\mu
+2))}{\mu^{2}}$. \newline For $P_{1+}$ the eigenvalues of the reduced 2D
system are $e_{2}(P_{1+})=3+\sqrt{\frac{3}{2}} (\mu+2), e_{3}(P_{1+})= 2
\left(  2+\sqrt{6}\right)  $. It is a source for $\mu>-2-\sqrt{6}$,
nonhyperbolic for $\mu=-2-\sqrt{6}$, saddle for $\mu<-2-\sqrt{6}$. \newline
For $P_{1-}$ the eigenvalues of the reduced 2D system are $e_{2}%
(P_{1-})=3-\sqrt{\frac{3}{2}} (\mu+2), e_{3}(P_{1-})= 2 \left(  2-\sqrt
{6}\right)  $. It is a sink for $\mu>\sqrt{6}-2$, nonhyperbolic for $\mu
=\sqrt{6}-2$, saddle for $\mu<\sqrt{6}-2$.

For $P_{3}$ the eigenvalues of the reduced 2D system are $e_{1}(P_{3})=\mu
(\mu+2)-2, e_{2}(P_{3})=\frac{1}{2} (\mu(\mu+4)-2).$ It is a sink for
$-1-\sqrt{3}<\mu<\sqrt{6}-2$, nonhyperbolic for $\mu=-1-\sqrt{3}$, or
$\mu=\sqrt{6}-2$, saddle otherwise. \newline For $P_{5}$, the eigenvalues are
$e_{1}(P_{5})=1, e_{2}(P_{5})=1-\frac{3 \mu}{2}$. It is a source for
$\mu<\frac{2}{3}$, nonhyperbolic for $\mu=\frac{2}{3}$, it is a saddle for
$\mu>\frac{2}{3}$. For $P_{6}$ the eigenvalues of the reduced system are
$e_{1}(P_{6})=-1+\frac{1}{\mu}-\frac{\sqrt{\mu\left(  6 \mu^{2}+9
\mu-22\right)  +9}}{\mu}, e_{2}(P_{6})=-1+\frac{1}{\mu}+\frac{\sqrt{\mu\left(
6 \mu^{2}+9 \mu-22\right)  +9}}{\mu}$. Nonhyperbolic for $\mu=-1-\sqrt{3}$,
saddle otherwise.

\begin{figure}[ptb]
\centering
\includegraphics[scale=0.6]{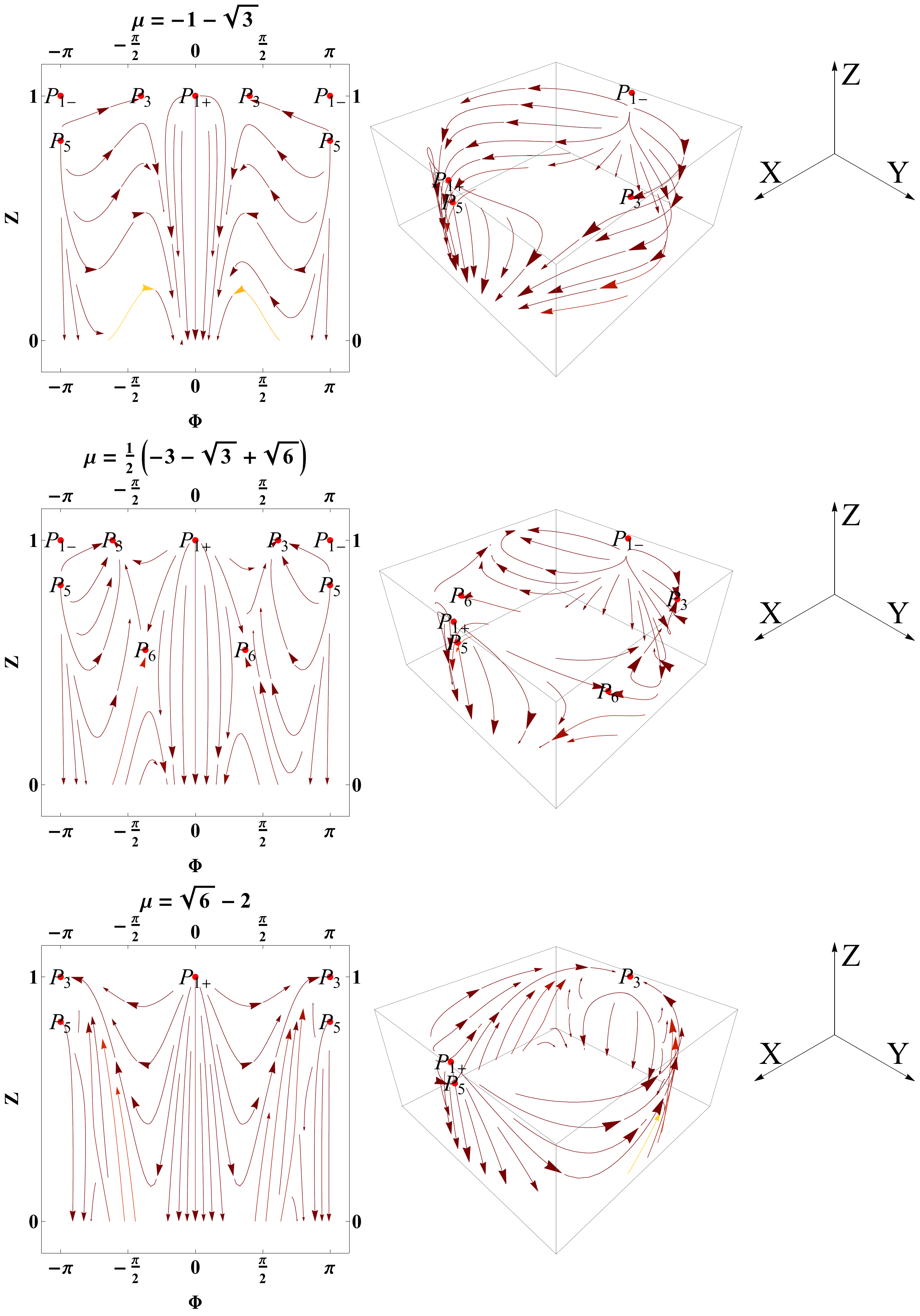} \caption{Unwrapped solution space
(left panel) and projection over the cylinder $\mathbf{S}$ (right panel) of
the solution space of system \eqref{eq47} - \eqref{eq48} for $\mu=-1-\sqrt{3},
\frac{1}{2}\left(  -3-\sqrt{3}-\sqrt{6}\right)  , -2+\sqrt{6}$. The region
$Z<0$ is nonphysical.}%
\label{fig:Plot1}%
\end{figure}

In Figure \ref{fig:Plot1} the unwrapped solution space (left panel) and
projection over the cylinder $\mathbf{S}$ (right panel) of the solution space
of system \eqref{eq47} - \eqref{eq48} for $\mu=-1-\sqrt{3}, \frac{1}{2}\left(
-3-\sqrt{3}-\sqrt{6}\right)  , -2+\sqrt{6}$ is presented. The domain of $\Phi$
was extended to $\Phi\in[-\pi,\pi]$ with ends $-\pi$ and $\pi$ identified.

\subsection{Analysis at infinity}

In figure \ref{fig:Plot1} are shown some orbits that approaches $Z=0$. The
region $Z<0$ is nonphysical. Combining \eqref{dyn.05} with \eqref{EQQ39} it
follows $r^{2}= x^{2}+y^{2}+\Sigma^{2} \rightarrow\infty$ as $Z\rightarrow0$.

Therefore, to analyze the dynamics at infinity the following variables $(\rho,
u, v)$ are defined:
\begin{equation}
\rho=\frac{r}{1+r}, \quad u=\tan^{-1}\left(  \frac{\Sigma}{x}\right)  , \quad
v=\tan^{-1}\left(  \frac{\sqrt{\Sigma^{2} + x^{2}}}{y}\right)  , \quad x\in(
-\infty, \infty), \quad y\in(0, \infty), \quad\Sigma\in( -\infty, \infty),
\end{equation}
such that, $\rho\rightarrow1$ when $r=\sqrt{x^{2}+y^{2}+ \Sigma^{2}}
\rightarrow\infty$. The inverse transformation is:
\begin{equation}
x=\frac{\rho}{1-\rho} \cos u\sin v,~\Sigma=\frac{\rho}{1-\rho}\sin u\sin
v,~y=\frac{\rho}{1-\rho}\cos v, \quad u \in\left[  0, 2\pi\right]  , \quad
v\in\left[  0,\frac{\pi}{2}\right]  .
\end{equation}

The solutions are drawn using the coordinates over the Poincar\'{e} sphere:
\begin{equation}
\bar{x}=\cos u\sin v,~\bar{\Sigma}=\sin u\sin v,~\bar{y}=\cos v.
\end{equation}

Introducing the time scaling
\begin{equation}
\frac{d f}{d T}= \frac{1}{1+r} \frac{d f}{d \lambda} \equiv(1-\rho)\frac{d
f}{d \lambda},
\end{equation}
the following dynamical system is obtained:
\begin{align}
&  \frac{d \rho}{d T}= \frac{1}{2} (2 \rho-1) \left(  \rho-2 \sqrt{2} (\rho-1)
\sin(v) \left(  \sin(u)+\sqrt{3} \cos(u)\right)  -3 \rho\cos(2 v)\right)  ,\\
&  \frac{d u}{d T}= \frac{\cos(u) \left(  \rho^{2} \cos(v) \cot(v) \left(
\sqrt{3} \mu\tan(u)+2\right)  +\left(  \sqrt{3} \tan(u)-1\right)  \csc(v)
\left(  (\rho-4) \rho+\rho^{2} \cos(2 v)+2\right)  \right)  }{\sqrt{2} \rho
},\\
&  \frac{d v}{d T}= -\frac{\cos(v) \left(  \sqrt{6} (\rho((\mu+2) \rho-4)+2)
\cos(u)+2 \sqrt{2} (1-2 \rho) \sin(u)-6 (\rho-1) \rho\sin(v)\right)  }{2 \rho
}.
\end{align}

As $\rho\rightarrow1^{-}$, the leading terms are:
\begin{align}
&  \frac{d \rho}{d T} = \frac{1}{2} (1-3 \cos(2 v))+\frac{1}{2} \left(  -9
\cos(2 v)-2 \sqrt{2} \left(  \sqrt{3} \cos(u)+\sin(u)\right)  \sin
(v)+3\right)  (\rho-1)+O\left(  (\rho-1)^{2}\right)  ,\\
&  \frac{d u}{d T} = \left(  \sqrt{2} \cos(u) \csc(v)+\sqrt{\frac{3}{2}}
\left(  \mu\csc^{2}(v)-\mu-2\right)  \sin(u) \sin(v)\right) \nonumber\\
&  +\left(  \sqrt{2} \cos(u) \csc(v)+\sqrt{\frac{3}{2}} \left(  \mu\csc
^{2}(v)-\mu-2\right)  \sin(u) \sin(v)\right)  (\rho-1)+O\left(  (\rho
-1)^{2}\right)  ,\\
&  \frac{d v}{d T} = \frac{\cos(v) \left(  2 \sin(u)-\sqrt{3} \mu
\cos(u)\right)  }{\sqrt{2}}+\frac{1}{2} \cos(v) \left(  -\sqrt{6} \mu\cos(u)+2
\sqrt{2} \sin(u)+6 \sin(v)\right)  (\rho-1)+O\left(  (\rho-1)^{2}\right)  .
\end{align}

In the limit $\rho\rightarrow1^{-}$, the radial equation becomes
\begin{equation}
\label{eqrho}\frac{d \rho}{d T}=\frac{1}{2} (1 - 3 \cos(2 v)),
\end{equation}
and it is independent of $\rho$. Therefore, the stationary points at infinity
are found by setting
\begin{subequations}
\label{compatibility}%
\begin{align}
&  \sqrt{2} \cos(u) \csc(v)+\sqrt{\frac{3}{2}} \left(  \mu\csc^{2}%
(v)-\mu-2\right)  \sin(u) \sin(v)=0,\\
&  \frac{\cos(v) \left(  2 \sin(u)-\sqrt{3} \mu\cos(u)\right)  }{\sqrt{2}}=0.
\end{align}
The stability of the stationary points at infinity is found as follows. First,
the stability of the pairs $(u^{*}, v^{*})$ which satisfy the compatibility
conditions \eqref{compatibility} are determined in the plane $u$--$v$. Then,
the global stability is examined by substituting in \eqref{eqrho} and
analyzing the sign of $\rho^{\prime*},v^{*})$. The sign $\rho^{\prime*}%
,v^{*})>0$ means that the region $\rho=1$ is approached meaning stability in
the radial coordinate, whereas $\rho^{\prime*},v^{*})<0$ means instability.

In table \ref{taab} is offered information about the location and existence
conditions of these critical points.

In figure \ref{figg1}, the dynamics of the stationary points at infinity of
system (\ref{dyn.02})-(\ref{dyn.04}) in the plane $(u,v)$ (left panels),
where $u$ is the horizontal axis and $v$ the vertical one, and over
the Poincar\`{e} sphere $\bar{x}=\cos u\sin v,~\bar{\Sigma}=\sin u\sin
v,~\bar{y}=\cos v$ (middle panels) are presented. The axis of the 3D
figures are drawn to the right.

\begin{table}[t]
\resizebox{\textwidth}{!}{     \centering
\begin{tabular}{|c|c|c|c|}\hline
Label $: \left(\begin{array}{c}
u  \\
v
\end{array}\right)$ & $\left(\begin{array}{c}
\bar{x} \\\bar{\Sigma} \\\bar{y}
\end{array}\right)$ & $\left(\begin{array}{c}
\rho'  \\
\lambda_1\\
\lambda_2
\end{array}\right)$ & Stability \\\hline\hline
$Q_1:\left(\begin{array}{c}
\pi  \left(2 c_1+\frac{1}{6}\right) \\
\frac{1}{2} \pi  \left(4 c_2+1\right) \\
\end{array}\right)$
&
$\left(\begin{array}{c}
\frac{\sqrt{3}}{2} \\
\frac{1}{2} \\
0 \\
\end{array}\right)$
&
$\left(\begin{array}{c}
2 \\
-2 \sqrt{2} \\
-\frac{2-3 \mu}{2 \sqrt{2}} \\
\end{array}\right)$ & $\begin{array}{c}
\text{sink for $\mu<\frac{2}{3}$}  \\
\text{saddle for $\mu>\frac{2}{3}$}
\end{array}$
\\
\hline
$Q_2: \left(\begin{array}{c}
\pi  \left(2 c_1+\frac{7}{6}\right) \\
\frac{1}{2} \pi  \left(4 c_2+1\right) \\
\end{array}\right)$
&
$\left(\begin{array}{c}
-\frac{\sqrt{3}}{2} \\
-\frac{1}{2} \\
0 \\
\end{array}\right)$
&
$\left(\begin{array}{c}
2 \\
2 \sqrt{2} \\
\frac{2-3 \mu }{2 \sqrt{2}} \\
\end{array}\right)$ & saddle
\\
\hline
$Q_3: \left(\begin{array}{c}
2 \pi  c_1+\tan ^{-1}\left(\frac{\sqrt{3} \mu }{2}\right) \\
2 \pi  c_2+\tan ^{-1}\left(\frac{\sqrt{3 \mu ^2+4}}{\sqrt{6 \mu -4}}\right) \\
\end{array}\right)$
&
$\left(\begin{array}{c}
\frac{2}{\sqrt{3} \sqrt{\mu  (\mu +2)}} \\
\frac{\mu }{\sqrt{\mu  (\mu +2)}} \\
\frac{\sqrt{2 \mu -\frac{4}{3}}}{\sqrt{\mu  (\mu +2)}} \\
\end{array}\right)$
&
$\left(\begin{array}{c}
2 \left(-\frac{4}{\mu +2}+\frac{1}{\mu }+1\right) \\
\frac{(\mu +2) \left(-\mu  \sqrt{3 \mu ^2+4} (\mu +2)-\sqrt{-\mu ^2 (\mu +2) \left(3 \mu ^2+4\right) (3 \mu  (4 \mu -3)-2)}\right)}{(\mu  (\mu
+2))^{3/2} \sqrt{6 \mu ^2+8}} \\
\frac{(\mu +2) \left(\sqrt{-\mu ^2 (\mu +2) \left(3 \mu ^2+4\right) (3 \mu  (4 \mu -3)-2)}-\mu  (\mu +2) \sqrt{3 \mu ^2+4}\right)}{(\mu  (\mu
+2))^{3/2} \sqrt{6 \mu ^2+8}} \\
\end{array}\right)$ &  $\begin{array}{c}
\text{saddle for $\mu<\frac{2}{3}$}  \\
\text{sink for $\mu>\frac{2}{3}$}
\end{array}$
\\\hline
$Q_4:\left(\begin{array}{c}
(2  c_1 +1)\pi +\tan ^{-1}\left(\frac{\sqrt{3} \mu }{2}\right) \\
2 \pi  c_2+\tan ^{-1}\left(\frac{\sqrt{3 \mu ^2+4}}{\sqrt{6 \mu -4}}\right) \\
\end{array}\right)$
&
$\left(\begin{array}{c}
-\frac{2}{\sqrt{3} \sqrt{\mu  (\mu +2)}} \\
-\frac{\mu }{\sqrt{\mu  (\mu +2)}} \\
\frac{\sqrt{2 \mu -\frac{4}{3}}}{\sqrt{\mu  (\mu +2)}} \\
\end{array}\right)$
&
$\left(\begin{array}{c}
2 \left(-\frac{4}{\mu +2}+\frac{1}{\mu }+1\right) \\
\frac{(\mu +2) \left(\mu  (\mu +2) \sqrt{3 \mu ^2+4}-\sqrt{-\mu ^2 (\mu +2) \left(3 \mu ^2+4\right) (3 \mu  (4 \mu -3)-2)}\right)}{(\mu  (\mu
+2))^{3/2} \sqrt{6 \mu ^2+8}} \\
\frac{(\mu +2) \left(\mu  \sqrt{3 \mu ^2+4} (\mu +2)+\sqrt{-\mu ^2 (\mu +2) \left(3 \mu ^2+4\right) (3 \mu  (4 \mu -3)-2)}\right)}{(\mu  (\mu
+2))^{3/2} \sqrt{6 \mu ^2+8}} \\
\end{array}\right)$ & saddle
\\
\hline
\end{tabular}}\caption{ Stability of the stationary points at infinity of
system (\ref{dyn.02})-(\ref{dyn.04}). $c_{1}$ and $c_{2}$ are integers. }%
\label{taab}%
\end{table}

\begin{figure}[th]
\centering
\includegraphics[scale=0.35]{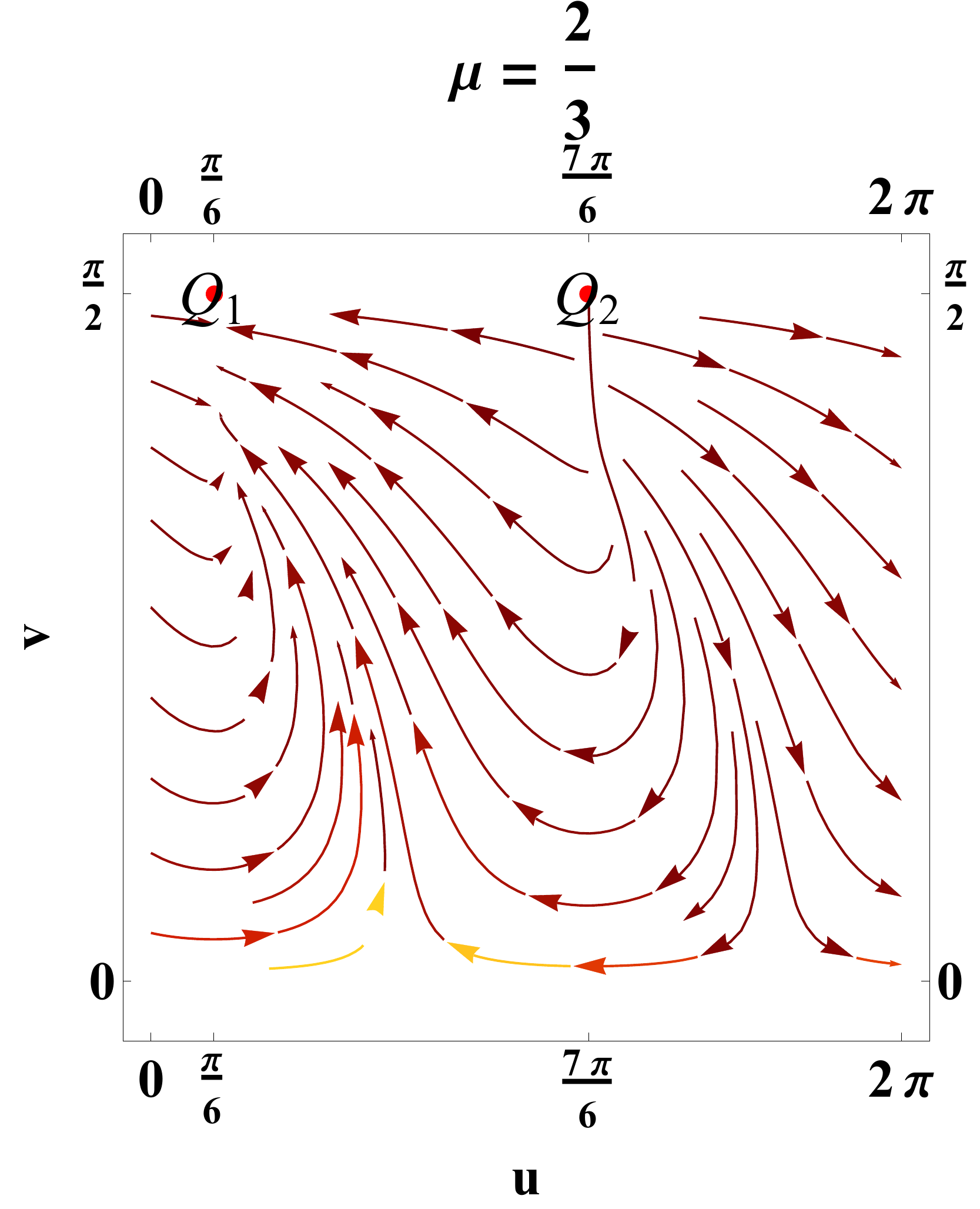} \includegraphics[scale=0.45]{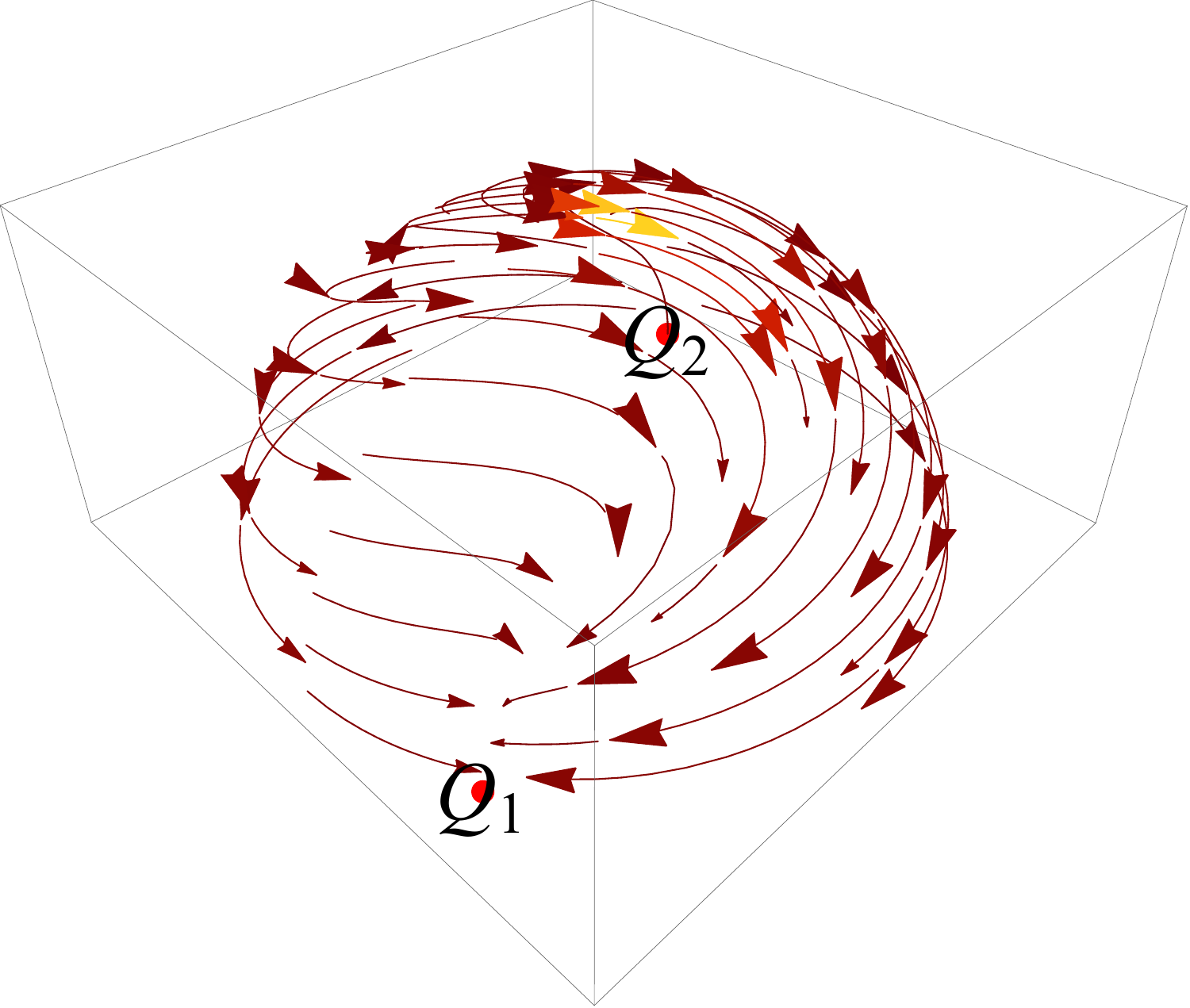}
\includegraphics[scale=0.35]{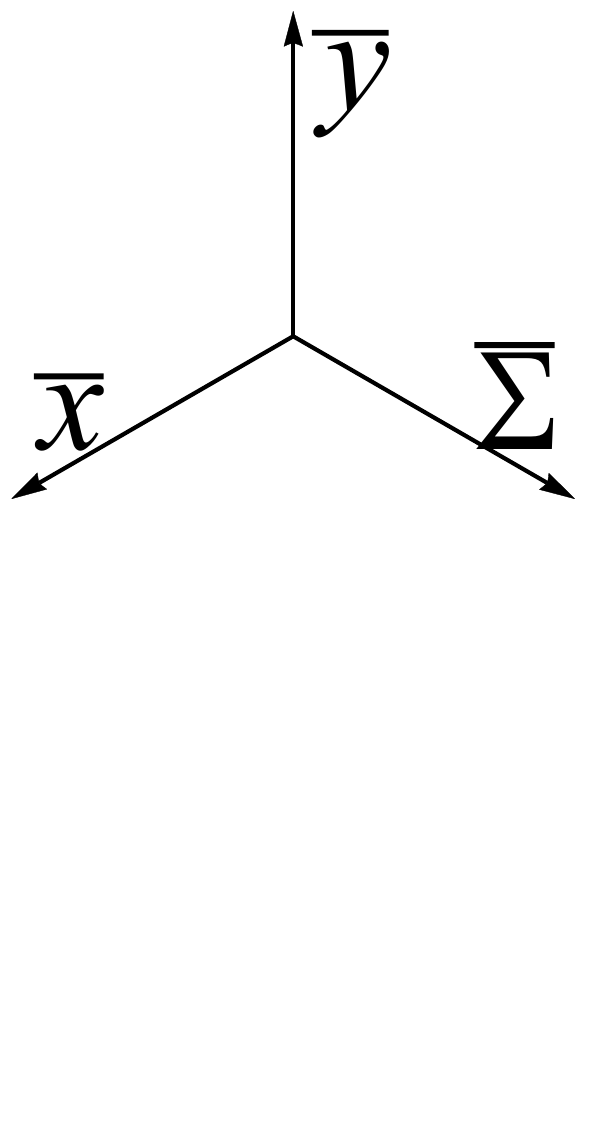} \includegraphics[scale=0.35]{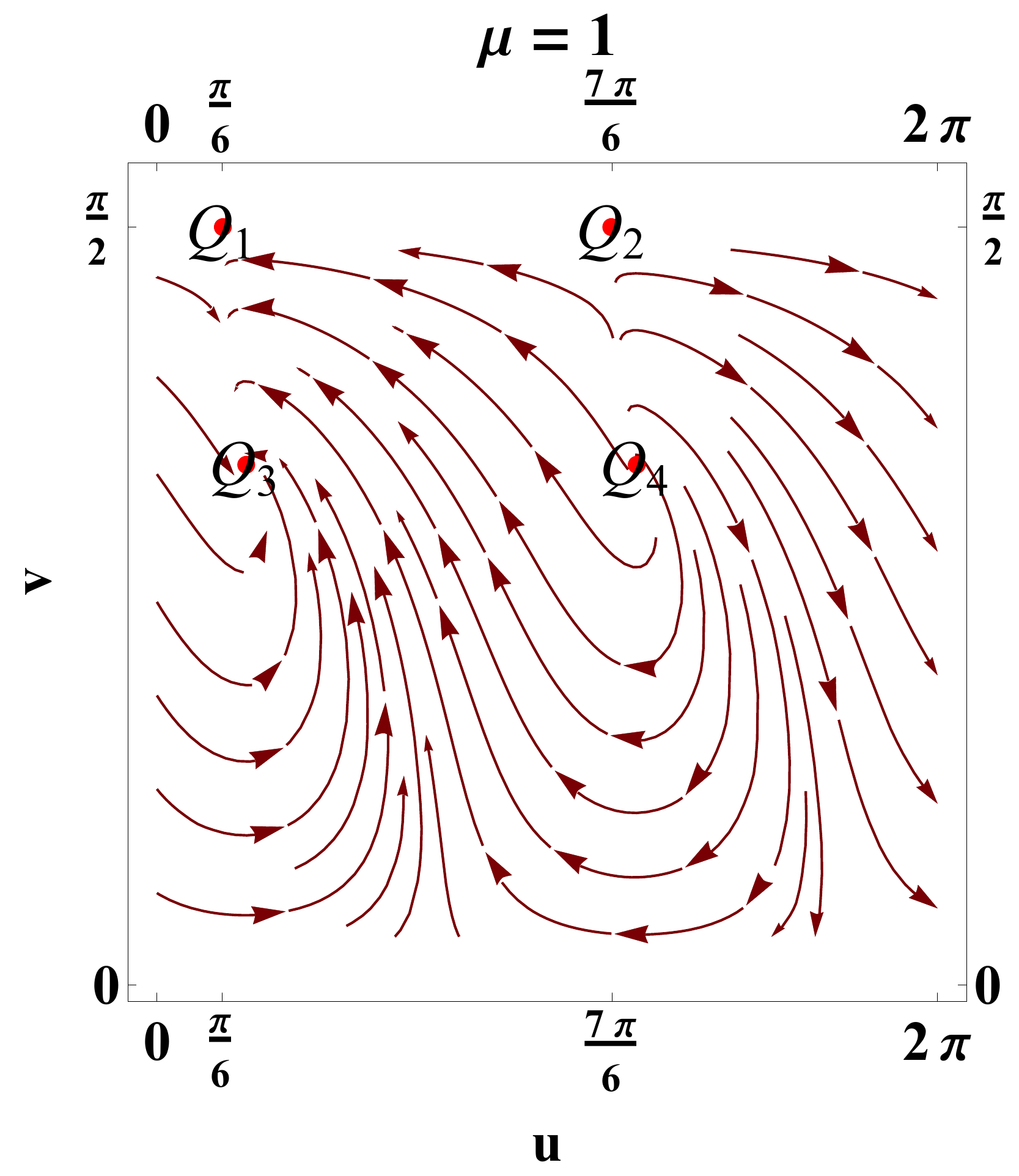}
\includegraphics[scale=0.45]{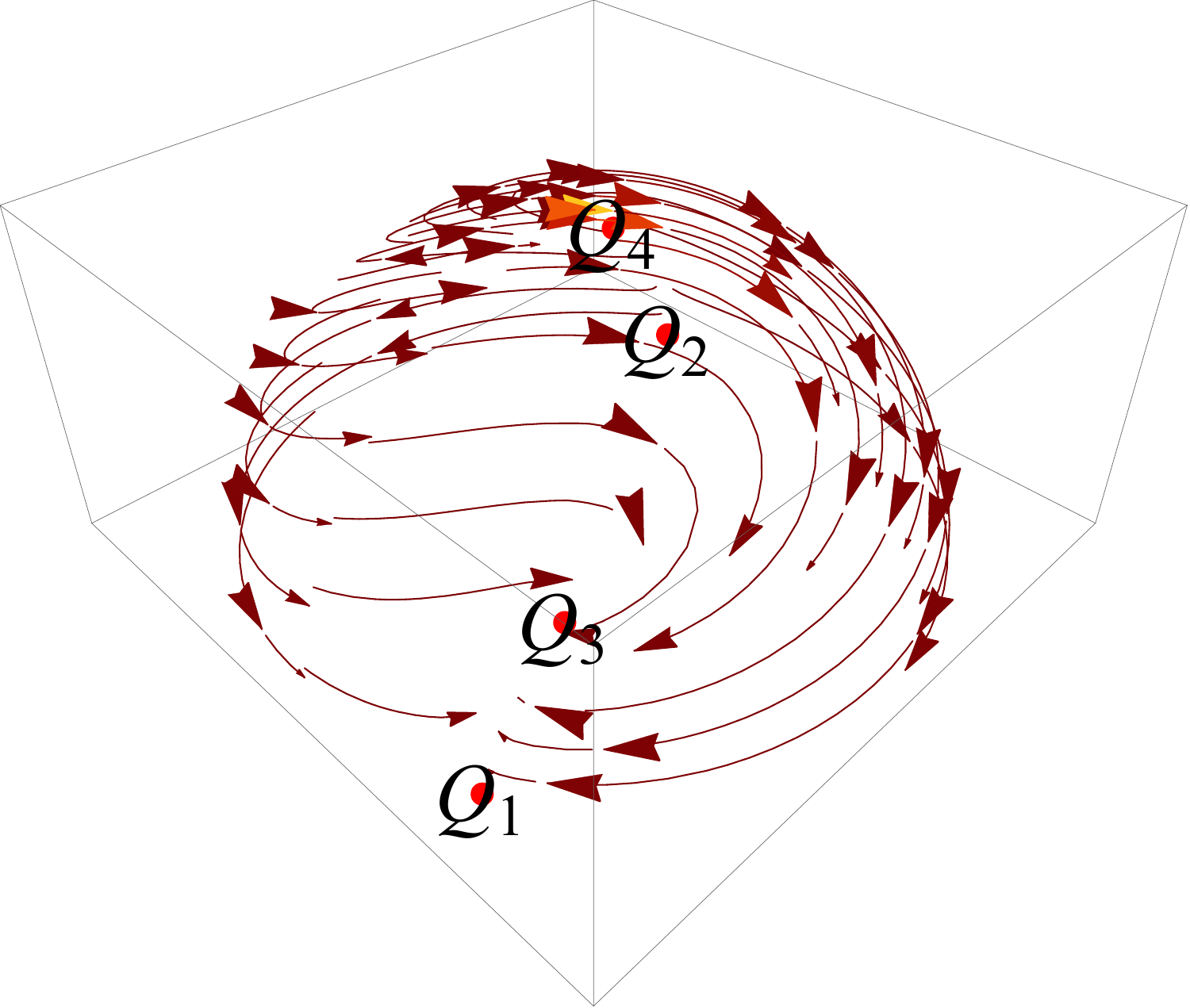} \includegraphics[scale=0.35]{axis}
\includegraphics[scale=0.35]{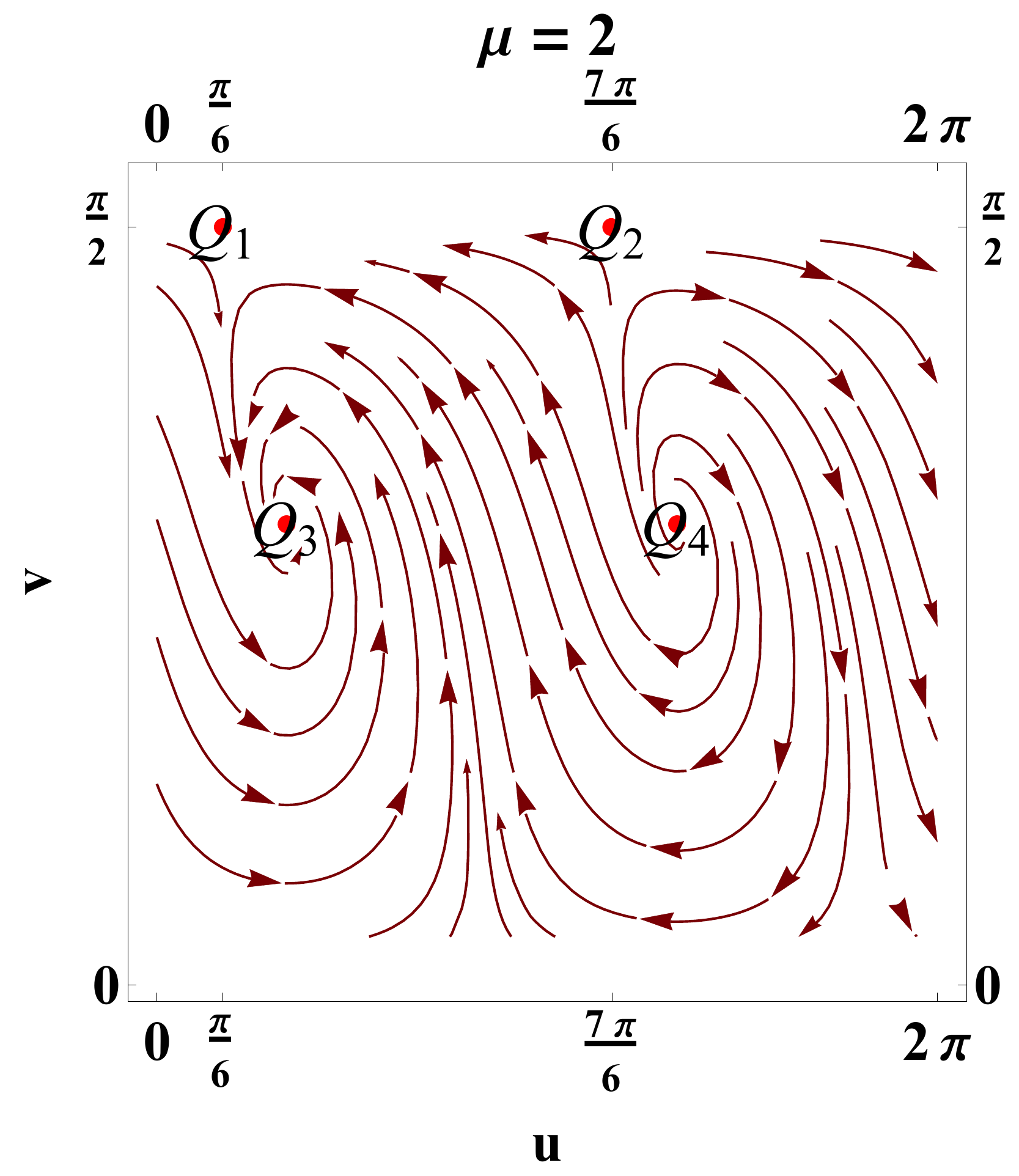} \includegraphics[scale=0.45]{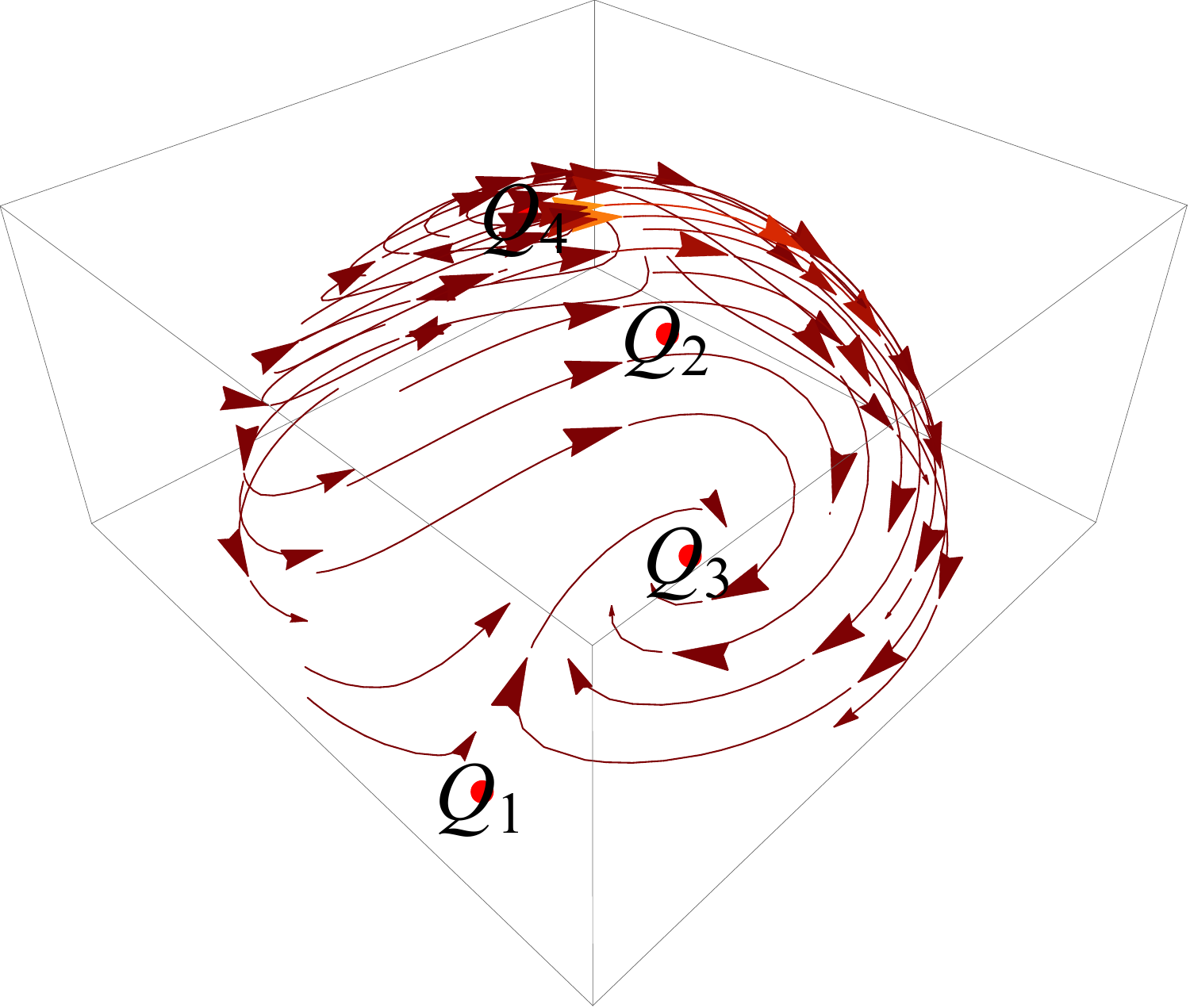}
\includegraphics[scale=0.35]{axis}\caption{ Dynamics of the stationary points
at infinity of system (\ref{dyn.02})-(\ref{dyn.04}) in the plane $(u,v)$ (left
panels) and the projection over the Poincar\`{e} sphere $\bar{x}=\cos u\sin
v,~\bar{\Sigma}=\sin u\sin v,~\bar{y}=\cos v$ (middle panels). }%
\label{figg1}%
\end{figure}

\section{Conclusions}

In this work in the context of Einstein-Scalar field theory, we determined
exact and analytic solutions for the gravitational field equations for a
Kantowski-Sachs background spacetime. The gravitational field equations
provides three unknown functions which are, the scalar field potential and the
coupling functions of the scalar field with the aether field.

For the determination of these unknown functions we apply a geometric
selection rule. In particular we require the existence of point
transformations which leave the field equations invariants, while from the
point transformations we can construct conservation laws, i.e. integrals of
motion, such that to simplify the nonlinear field equations and determine the
exact solutions.

In addition, the asymptotic behaviour of the field equations is studied, from
where we find that the limits of Bianchi I and closed FLRW spacetimes exist.
In order to perform a complete and detailed analysis on the determination of
the stationary points we work with two different sets of dimensionless
variables, the $H$-normalization approach, and the compactification procedure.
We observe that the second set of dimensionless variables provides additional
information for the evolution of the dynamical system.

In a future work we plan to study by using this approach the case of static
spherically symmetric spacetimes and study the existence of black-holes solutions.
\end{subequations}
\begin{acknowledgments}
AP\ \& GL were funded by Agencia Nacional de Investigaci\'{o}n y Desarrollo -
ANID through the program FONDECYT Iniciaci\'{o}n grant no. 11180126.
Additionally, GL is supported by Vicerrector\'{\i}a de Investigaci\'{o}n y
Desarrollo Tecnol\'{o}gico at Universidad Catolica del Norte.
\end{acknowledgments}


\begin{thebibliography}{99}                                                                                               %


\bibitem {Mis69}\ C.W. Misner, The Isotropy of the universe, Ap. J.
\textbf{151}, 431 (1968)

\bibitem {szydl}O. Hrycyna and M. Szydlowski, AIP Conf.Proc. \textbf{1514},
191 (2013)

\bibitem {russ}E. Russel, C. Battal Kilinc and O.K Pashaev, MNRAS
\textbf{442}, 2331 (2014)

\bibitem {guth}A. Guth, Phys. Rev. D \textbf{23}, 347 (1981)

\bibitem {b1}A. Feinstein and J. Ibanez, Class. Quantum Grav. \textbf{10}, 93 (1993)

\bibitem {b2}J.M. Aguirregabiria, A. Feinstein and J. Ibanez, Phys. Rev. D
\textbf{48}, 4662 (1993)

\bibitem {b3}M. Tsamparlis and A. Paliathanasis, Gen. Relativ. Gravit.
\textbf{43}, 1861 (2010)

\bibitem {b4}T. Christodoulakis, G. Kofinas, E. Korfiatis, G.O. Papadopoulos
and A. Paschos, J. Math. Phys. \textbf{42}, 3580 (2001)

\bibitem {b5}T. Christodoulakis and P. A. Terzis, Class. Quantum Grav.
\textbf{24}, 875 (2007)

\bibitem {b6}J. Socorro, L.O. Pimenter, C. Ortiz and M. Aguero, Int. J. Theor.
Phys. \textbf{48}, 3567 (2009)

\bibitem {b7}M. Thorsud, Class. Quantum Grav. \textbf{36}, 235014 (2019)

\bibitem {b8}T. Christodoulakis, T. Grammenos, C. Helias, P.G. Kevrekidis and
A. Spanou, J. Math. Phys. \textbf{47}, 042505 (2006)

\bibitem {kanno}S. Kanno and J. Soda, Phys. Rev. D \textbf{74}, 063505 (2006)

\bibitem {DJ}W.~Donnelly and T.~Jacobson, Phys. Rev. D \textbf{82}, 064032 (2010)

\bibitem {eas1}J.D.~Barrow, Phys. Rev. D \textbf{85}, 047503 (2012)

\bibitem {eas2}A. Paliathanasis, G. Papagiannopoulos, S. Basilakos and J.D.
Barrow, EPJC \textbf{79}, 723 (2019)

\bibitem {ea1}A. Paliathanasis and G. Leon, EPJC \textbf{80}, 355 (2020)

\bibitem {ea2}A. Paliathanasis and G. Leon, EPJC \textbf{80}, 589 (2020)

\bibitem {KS1}R. Kantowski and R. K. Sachs, J. Math. Phys. \textbf{7}, 443 (1966)

\bibitem {nns1}M. Thorsrud, B.D. Normann and\ T.S. Pereira, Class. Quantum
Grav. \textbf{37}, 065015 (2020)

\bibitem {Byland:1998gx}S.~Byland and D.~Scialom, Phys. Rev. D \textbf{57},
6065-6074 (1998)

\bibitem {Fadragas:2013ina}C.~R.~Fadragas, G.~Leon and E.~N.~Saridakis, Class.
Quant. Grav. \textbf{31}, 075018 (2014)

\bibitem {Latta:2016jix}J.~Latta, G.~Leon and A.~Paliathanasis, JCAP
\textbf{1611}, 051 (2016)

\bibitem {Coley:2015qqa}A.~A.~Coley, G.~Leon, P.~Sandin and J.~Latta, JCAP
\textbf{12}, 010 (2015)

\bibitem {mech3}N. Dimakis, A. Karagiorgos, A. Zampelis, A. Paliathanasis, T.
Christodoulakis and P.A. Terzis, Phys. Rev. D \textbf{93}, 123518 (2016)

\bibitem {mech4}N. Dimakis, P.A. Terzis and T. Christodoulakis, Phys. Rev. D
\textbf{99}, 104061 (2019)

\bibitem {eaqm}N. Dimakis, T. Pailas, A. Paliathanasis, G. Leon, P.A.\ Terzis
and T. Christodoulakis, [arXiv:2008.00746]

\bibitem {qm1}T. Christodoulakis, N. Dimakis, P.A. Terzis, G. Doulis, Th.
Grammeos, E. Melas and A. Spanou, J. Geom. Phys. \textbf{71}, 127 (2013)

\bibitem {qm2}T. Christodoulakis, N. Dimakis, P.A. Terzis and G. Doulis,
Phys.\ Rev. D \textbf{90}, 024052 (2014)

\bibitem {mech1}G. Papagiannopoulos, J.D. Barrow, S. Basilakos, A. Giacomini
and A. Paliathanasis, Phys.\ Rev. D \textbf{95}, 024021 (2017)

\bibitem {mech2}A. Paliathanasis, M.\ Tsamparlis, S. Basilakos and J.D.
Barrow, Phys.\ Rev. D \textbf{91}, 123535 (2015)

\bibitem {ns1}H. Motavali and M. Golshani, IJMPA \textbf{17}, 375 (2002)

\bibitem {ns2}U. Camci and Y. Kucukakca, Phys. Rev. D \textbf{76}, 084023 (2007)

\bibitem {ns3}J. Mubasher, F.M. Mahomed and D. Momeni, Phys. Lett. B
\textbf{702}, 315 (2011)

\bibitem {ns4}Y. Zhang, Y.-G. Gong, Z.-H. Zhu, Phys. Lett. B \textbf{688}, 13 (2010)

\bibitem {ns5}H. M. Sadjadi, Phys. Lett. B \textbf{718}, 270 (2012)

\bibitem {ns6}B. Vakili, F. Khazaie, Class. Quant. Grav. \textbf{29}, 035015 (2012)

\bibitem {ns7}B. Modak, S. Kamilya and S. Biswas, Gen. Relativ. Gravit.
\textbf{32}, 1615 (2000)

\bibitem {ns8}A. Paliathanasis, M. Tsamparlis and S. Basilakos, Phys. Rev. D
\textbf{84}, 123514 (2011)

\bibitem {ns9}J.A. Belinchon, T. Harko and M.K. Mak, Astroph. Sp. Sci.
\textbf{361}, 52 (2016)

\bibitem {Collins:1977fg}C.~B.~Collins, J. Math. Phys. \textbf{18}, 2116 (1977)

\bibitem {Weber:1984xh}E.~Weber, J. Math. Phys. \textbf{25}, 3279 (1984)

\bibitem {Gron:1986ua}O.~Gron, J. Math. Phys. \textbf{27}, 1490 (1986)

\bibitem {LorenzPetzold:1985jm}D.~Lorenz-Petzold, Phys. Lett. B \textbf{149},
79 (1984)

\bibitem {Solomons:2001ef}D.~M.~Solomons, P.~Dunsby and G.~Ellis, Class.
Quant. Grav. \textbf{23}, 6585 (2006)

\bibitem {Jamal:2017cut}S.~Jamal and G.~Shabbir, Eur. Phys. J. Plus
\textbf{132}, 70 (2017)

\bibitem {Zubair:2016ccy}M.~Zubair and S.~M.~Ali Hassan, Astrophys. Space Sci.
\textbf{361}, 149 (2016)

\bibitem {Alvarenga:2015jaa}F.~G.~Alvarenga, R.~Fracalossi, R.~C.~Freitas and
S.~V.~B.~Gon\c{c}alves, Braz. J. Phys. \textbf{48}, 370 (2018)

\bibitem {Barrow:1996gx}J.~D.~Barrow and M.~P.~Dabrowski, Phys. Rev. D
\textbf{55}, 630 (1997)

\bibitem {Camci:2016yed}U.~Camci, A.~Yildirim and I.~Basaran Oz, Astropart.
Phys. \textbf{76}. 29 (2016)

\bibitem {Barrow:2018zav}J.~D.~Barrow and A.~Paliathanasis, Eur. Phys. J. C
\textbf{78}, 767 (2018)

\bibitem {Calogero:2009mi}S.~Calogero and J.~M.~Heinzle, Physica D
\textbf{240}, 636 (2011)

\bibitem {Carr:1999qr}B.~J.~Carr and A.~A.~Coley, Phys. Rev. D \textbf{62},
044023 (2000)

\bibitem {deCesare:2020swb}M.~de Cesare, S.~S.~Seahra and E.~Wilson-Ewing,
JCAP 07, 018 (2020)

\bibitem {Clancy:1998ka}D.~Clancy, J.~E.~Lidsey and R.~K.~Tavakol, Class.
Quant. Grav. \textbf{15}, 257 (1998)

\bibitem {in1}R. Chan, M.F.A. da Silva and V.H. Satheeshkumar, The General
Spherically Symmetric Static Solutions in the Einstein-Aether Theory, [arXiv:2003.00227]

\bibitem {col1}B. Alhulaimi, R.J. van den Hoogen and A.A. Coley, JCAP
\textbf{17}, 045 (2017)

\bibitem {col2}R.J. van de Hoogen, A.A. Coley, B. Alhulaimi, S. Mohandas, E.
Knighton and S. O'Neil, JCAP \textbf{18}, 017 (2018)

\bibitem {Coley:2019tyx}A.~Coley and G.~Leon, Gen.\ Rel.\ Grav.\ \textbf{51},
no. 9, 115 (2019)

\bibitem {Leon:2019jnu}G.~Leon, A.~Coley and A.~Paliathanasis, Annals
Phys.\ \textbf{412}, 168002 (2020)

\bibitem {sbns}S. Basilakos, M. Tsamparlis and A. Paliathanasis, Phys. Rev. D
\textbf{83}, 103512 (2011)

\bibitem {ovv}L.V. Ovsiannikov, Group Analysis of Differential Equations,
Academic Press, New York (1982)

\bibitem {mtssym}M. Tsamparlis and A. Paliathanasis, Symmetry \textbf{10}, 233 (2018)
\end{thebibliography}
\end{document}